\input harvmac
\def \II {{\cal I}}
\def \tr {{\rm tr}}

\def \G {\Gamma}
\def \m {\mu}
\def \n {\nu}
\def \bK {{\bar K}}

\def   \W {{\rm W}}

\def \eight{{\textstyle{1\over 8}}}
\def \ha{{\textstyle{1\over 2}}}
\def \four {{\textstyle{1\over 4}}}
\def \K {{\cal K}}
\def \bbK {\bar {\cal K}}
\def \tF {\tilde F}
\def \exw {\langle {\W} \rangle}

\def \da {{\dot \a}}
\def \sF {{F^*}}
\def \ads {$AdS_5\times S^5\ $}
\def \T {{\cal T}}
\def \tr {{\rm tr}}
\def \Str {{\rm Str}}

\def \te {\textstyle}
\def \Str {{\rm Str}}

\def \W {{\cal W}}

\def \a {\alpha}
\def \ad {{\dot{\alpha}}}

\def \four {{1\ov 4}}

\def \ep{\epsilon}

\def \L {\Lambda}
\def \LL  {{\cal L}}

\def \N {{\cal N}}

\def \x {\xi} 

\def \bD {\bar{D}}
\def \BI {Born-Infeld }
\def \bD {\bar{D}}
\def \bW {\bar{W}}
\def \bP {\bar{\Phi}}
\def \bvp {{\bar{\vp}}}
\def \k {\kappa} 
\def \cF {{\cal F}}

\def \del {\partial}
\def \ada {{\a\da}}
\def \da {{\dot \a}}

\def \const {{\rm const}}

\def \b {\beta}

\def \s {\sigma}
\def \p {\phi}
\def \m {\mu}
\def \n {\nu}
\def \vp {\varphi }
\def \t {\theta}
\def \bt {\bar{\theta}}

\def \td {\tilde }
\def \d {\delta}

\def \P {\Phi}

\def \inv {^{-1}}
\def \ov {\over }
\def \four{{\textstyle{1\over 4}}}

\def \f {{\rm F}}
\def \N {{\cal N}}

\def \NO {$\N=1$ }
\def \NT {$\N=2$ }
\def \NF {$\N=4$ }
\def \t {\theta}

\def \l {\lambda}

\def \lr { \lref}
\def\np {{  Nucl. Phys. }}
\def \pl {{  Phys. Lett. }}
\def \mpl {{ Mod. Phys. Lett. }}
\def \pr  {{ Phys. Rev. }}

\def \prl {{  Phys. Rev.  Lett. }}

\gdef \jnl#1, #2, #3, 1#4#5#6{ { #1~}{ #2} (1#4#5#6) #3}

\lr\roch{F. Gonzalez-Rey, B. Kulik, I.Y. Park  and  M. Ro\v cek,
``Self-Dual Effective Action of $\N=4$ Super-Yang Mills",
 hep-th/9810152.}

\lr \bggg{ J.~Bagger and A.~Galperin, {``The Tensor Goldstone
multiplet for partially broken supersymmetry"},
Phys. Lett. B412 (1997) 296, hep-th/9707061.}

\lr\gonr{F. Gonzalez-Rey  and  M. Ro\v cek,
``Nonholomorphic $\N=2$ terms in $\N=4$ SYM: 1-Loop Calculation in 
$\N=2$ superspace", 
Phys. Lett. B434 (1998) 303, 
hep-th/9804010.}

\lr \peri{V. Periwal and R. von Unge, 
``Accelerating D-branes'', 
Phys. Lett. B430 (1998) 71, 
hep-th/9801121.}

\lr\gross {D.J.  Gross and E. Witten,
``Superstring modification of Einstein equations", 
 \np B277 (1986)  1.}

\lr \wiit{
 E. Witten, ``Anti De Sitter Space And Holography", 
Adv.Theor.Math.Phys. 2 (1998) 253, hep-th/9802150.}

\lr \apt{I. Antoniadis, H. Partouche and T.R. Taylor,
{``Spontaneous breaking of  $\N=2$ global supersymmetry"},
\pl B372 (1996) 83,
hep-th/9512006.}

\lr\gonz{F. Gonzalez-Rey, unpublished.}
\lr \howest{G.G. Harwell  and P.S. Howe, ``A superspace survey",
Class. Quant. Grav. 12 (1995) 1823;
P.S. Howe and P.C.  West,
 ``Operator product expansions in 
four-dimensional superconformal field theories'', 
Phys. Lett. B389 (1996) 273, hep-th/9607060.}

\lr \napp {P.C. Argyres and C.R. Nappi, 
``Spin 1 effective actions from open strings", 
\np B330 (1990) 151. }

 \lr \shif { M.A. Shifman, ``Wilson loop in vacuum fields",
 \np B173 (1980) 13.}
\lr \hasht {A. Hashimoto  and  W. Taylor,
``Fluctuation Spectra of Tilted and Intersecting D-branes from the Born-Infeld Action", 
Nucl.Phys. B503 (1997) 193, 
  hep-th/9703217.}

       \lr \martin{M. Ro\v cek, {``Linearizing the Volkov-Akulov model"},
\prl 41 (1978) 451.}

\lr \ttts{
 A.A. Tseytlin,  ``Open superstring partition function in constant gauge field background at finite temperature", 
 Nucl. Phys. B524 (1998) 41, 
 hep-th/9802133. 
}

 \lr \chep{
I. Chepelev  and A.A. Tseytlin,
 Long-distance interactions of branes: correspondence between supergravity and super Yang-Mills descriptions,
 Nucl. Phys. B515 (1998) 73, hep-th/9709087.}

\lr \maldstr{
J.M. Maldacena,  ``Branes probing black holes", 
Nucl.Phys.Proc.Suppl. 68 (1998) 17,
hep-th/9709099.}

\lr \malda{
J.M.  Maldacena,
``The large N limit of superconformal field theories
and supergravity", Adv.Theor.Math.Phys. 2 (1998) 231, 
hep-th/9711200.}

\lr\keski{
E. Keski-Vakkuri and P. Kraus, ``Born-Infeld actions from matrix theory",
\np B518 (1998) 212, hep-th/9709122;
S. de Alwis, ``Matrix models and string world sheet duality",
\pl B423 (1998) 59, hep-th/9710219;
V. Balasubramanian, R. Gopakumar  and  F. Larsen,
``Gauge theory, geometry and the large N limit",
hep-th/9712077.}

\lr \kalo{R. Kallosh, J. Kumar and  A. Rajaraman,
``Special conformal symmetry of world-volume actions",
 Phys. Rev. D57  (1998) 6452,
 hep-th/9712073;
P. Claus, R. Kallosh, J. Kumar, P. Townsend  and  A. van Proeyen,
``Conformal Theory of M2, D3, M5 and D1-Branes+D5-branes",
JHEP 9806 (1998) 004, 
hep-th/9801206.}

\lr\tayl{ W. Taylor  and  M. Van Raamsdonk,
  ``Multiple D0-branes in Weakly Curved Backgrounds", hep-th/9904095.}

\lr\deser{ S. Deser, J. McCarthy  and  O. Sarioglu,
``Good Propagation Constraints on Dual Invariant
Actions in Electrodynamics and on Massless Fields", 
 Class.Quant.Grav. 16 (1999) 841, hep-th/9809153.}

\lr \odint{S. Nojiri  and S.D. Odintsov,
``On the instability of effective potential for non-abelian
toroidal D-brane", 
Phys. Lett. B419 (1998) 107,   hep-th/9710137.}

\lr \born { M. Born, {Proc. Roy. Soc. } {A 143} (1934) 410; 
 M. Born and L. Infeld, ``Foundations of the new field theory",
{Proc. Roy. Soc. } {A 144} (1934) 425;
M. Born, ``Th\'eorie non-lin\'eare du champ \'electromagn\'etique", 
{Ann. Inst. Poincar\'e}, 7 (1939) 155.  } 

\lr \polyak{Polyakov   Wall of the cave }

\lr \mt{
R.R. Metsaev and A.A. Tseytlin,
``Type IIB superstring action in $AdS_5 \times S^5$ background",
hep-th/9805028.}

\lr\jhs{M. Aganagic, C. Popescu and J.H. Schwarz, hep-th/9612080;
E. Bergshoeff and P.K. Townsend, hep-th/9611173.}

\lr\bst {E. Bergshoeff, E. Sezgin and P.K. Townsend, \pl B189 (1988) 75.  }
\lr \ffff{E. Bergshoeff, M. de Roo and T. Ort\'in, hep-th/9606118;
M. Aganagic, J. Park, C. Popescu and J.H. Schwarz, hep-th/9701166;
I. Bandos, K. Lechner, A. Nurmagambetov, P. Pasti, D. Sorokin and M. Tonin, 
hep-th/9701149.} 

\lr \ferr{S. Ferrara, M.A. Lledo and A. Zaffaroni, 
 ``Born-Infeld Corrections to D3 brane Action in $AdS_5\times S_5$ and $\N=4$, d=4
     Primary Superfields'', 
 Phys. Rev. D58 (1998) 105029, 
hep-th/9805082.}

\lr \aha{O. Aharony, S.S. Gubser, J. Maldacena, H. Ooguri  and 
 Y. Oz, ``Large N field theories, string theory and gravity",  
 hep-th/9905111.} 

\lr \don{  J.F. Donoghue, ``Introduction to the Effective Field Theory Description of Gravity", gr-qc/9512024.
 }

\lr \gsw{R. Grimm, M. Sohnius and J. Wess,
{``Extended supersymmetry and gauge theories"},
\np B133 (1978) 275. }

\lr\klebb{
S.S. Gubser, I.R.  Klebanov and A.A.
  Tseytlin, ``String theory and classical
  absorption by three-branes", 
  \np B499 (1997) 41, hep-th/9703040.}
\lr \kleb { I.R. Klebanov, ``World-volume approach
to absorption by non-dilatonic branes",
\np B496 (1997) 231, hep-th/9702076.}
\lr \howest{G.G. Harwell  and P.S. Howe, ``A superspace survey",
Class. Quant. Grav. 12 (1995) 1823;
P.S. Howe and P.C.  West,
 ``Operator product expansions in 
four-dimensional superconformal field theories'', 
Phys. Lett. B389 (1996) 273, hep-th/9607060.}
\lr \eden{B. Eden, P.S. Howe and P.C. West, `` Nilpotent invariants in  $\N=4$ SYM'',
hep-th/9905085.}
\lr \guys { O. Aharony, S. S. Gubser, J. Maldacena, H. Ooguri  and  Y. Oz,
``Large N Field Theories, String Theory and Gravity'', 
hep-th/9905111.}

\lr \meets{
 R.R. Metsaev  and  A.A. Tseytlin, ``Supersymmetric D3-brane action in 
$AdS_5  \times  S^5$'', 
 Phys. Lett. B436 (1998) 281,  
 hep-th/9806095.}

\lr \mrt{
 R.R. Metsaev, M.A. Rakhmanov  and  A.A. Tseytlin, ``The Born-Infeld
     action as the effective action in the open superstring theory'', 
    Phys. Lett. B193 (1987) 207.
 }

\lr\sch{ 
M. Cederwall, A. von Gussich, B.E.W. Nilsson and A. Westenberg,
The Dirichlet super-three-brane in ten-dimensional type IIB supergravity,
Nucl. Phys. {B}490 (1997) 163, hep-th/9610148; 
M. Aganagic, C. Popescu and J.H. Schwarz, D-brane actions with local kappa-symmetry,
Phys. Lett. {B}393 (1997) 311, hep-th/9610249;
Gauge-invariant and and gauge-fixed D-brane actions,
Nucl. Phys. {B}495 (1997) 99, hep-th/9612080; 
E. Bergshoeff and  P.K. Townsend, Super D-branes,
Nucl. Phys. {B}490 (1997) 145,
hep-th/9611173.
}
\lr\roche{  F. Gonzalez-Rey, I.Y. Park  and  M. Ro\v cek, 
``On dual 3-brane actions with partially broken  $\N=2$ supersymmetry'',
 Nucl. Phys. B544  (1999) 243, 
 hep-th/9811130.
   } 
\lr \bag {J. Bagger and A. Galperin, {``Matter couplings in
partially broken extended supersymmetry"},
\pl B336 (1994) 25, hep-th/9406217.} 

\lr \rot  { M. Ro\v cek and A.A. Tseytlin, ``Partial breaking of global 
D=4 supersymmetry, constrained superfields, and 3-brane actions'',
Phys. Rev. D59 (1999) 106001,  hep-th/9811232.}

\lr \ktv{ I. Klebanov, W. Taylor  and  M. Van Raamsdonk,
``Absorption of dilaton partial waves by D3-branes'', 
hep-th/9905174.}

\lr \niel{H.B. Nielsen and P. Olesen, \np B57 (1973) 367;
 ``Vortex line models for dual strings", \np B61 (1973) 45;
H.C. Tze, ``Born duality and strings in hadrodynamics and electrodynamics", 
Nuov. Cim. 22A (1974) 507.}

\lr \des{S. Deser and R. Puzalowski, {``Supersymmetric nonpolynomial
vector multiplets and causal propagation"},
J. Phys. A13 (1980) 2501.}

\lr \berg{E. Bergshoeff, E. Sezgin, C.N. Pope  and  P.K. Townsend,
``The Born-Infeld action from conformal invariance 
of the open superstring", 
 Phys. Lett. B188  (1987) 70. 
 }

\lr\moun{ J.A. de Azcarraga, A.J. Macfarlane, A.J. Mountain and 
 J.C. Perez Bueno, ``Invariant tensors for simple groups",
 Nucl. Phys. B510 (1998) 657, 
 physics/9706006;
A.J. Mountain,
``Invariant tensors and Casimir operators for simple compact Lie groups", 
 J. Math. Phys. 39 (1998) 5601,
physics/9802012.}

\lr \bergr{
 E. Bergshoeff, M. Rakowski  and  E. Sezgin,
``Higher derivative super Yang-Mills theories",
Phys. Lett. B185 (1987) 371. }

\lr \ket{ S.V. Ketov,
`` A manifestly  $\N=2$ supersymmetric Born-Infeld action",
 Mod. Phys. Lett. A14 (1999) 501, 
  hep-th/9809121. }

\lr \apt{I. Antoniadis, H. Partouche and T.R. Taylor,
{``Spontaneous breaking of  $\N=2$ global supersymmetry"},
\pl B372 (1996) 83,
hep-th/9512006.}

\lr \calm{ C.G. Callan and J. Maldacena, ``Brane dynamics from the 
Born-Infeld action", Nucl. Phys. B513 (1998) 198,
hep-th/9708147. }

\lr \gibb { G.W. Gibbons, ``Born-Infeld particles and Dirichlet p-branes", 
Nucl. Phys. B514 (1998) 603, hep-th/9709027. }

\lr \thorl{L. Thorlacius, ``Born-Infeld string as a 
boundary conformal field theory", Phys. Rev. Lett. 80 (1998) 1588,
 hep-th/9710181.}

\lr \gibbo{G. Gibbons and A.A. Tseytlin, unpublished (1997).}

\lr  \witt { E. Witten,
``Bound states of strings and $p$-branes", 
 \np B443 (1995) 85, hep-th/9510135. } 

\lr \bac { C. Bachas, ``D-brane dynamics",
 \pl B374 (1996) 37, hep-th/9511043. }

\lr\gonor{ S. Gonorazky, F.A. Schaposnik  and  G. Silva, 
``Supersymmetric non Abelian Born-Infeld theory",
 Phys. Lett.  B449 (1999) 187,   
hep-th/9812094.
}

\lr\gon{N. Grandi, R.L. Pakman, F.A. Schaposnik and  G. Silva, 
``Monopoles, dyons and theta term in Dirac-Born-Infeld theory",  hep-th/9906244;
N. Grandi, E.F. Moreno  and  F.A. Schaposnik, 
``Monopoles in non Abelian Dirac-Born-Infeld theory", 
 Phys.  Rev. D59 (1999) 125014, hep-th/9901073;
 P.K. Tripathy, 
``Gravitating monopoles and black holes in
Einstein-Born-Infeld-Higgs model", 
hep-th/9904186, 
 Phys. Lett. B458 (1999) 252.
}
\lr \nun{H.R. Christiansen, C. Nunez and  F.A. Schaposnik, 
``Uniqueness of Bogomol'nyi equations and Born-Infeld like supersymmetric theories",  
 Phys. Lett.  B441 (1998) 185, 
 hep-th/9807197; 
S. Gonorazky, C. Nunez, F.A. Schaposnik  and  G. Silva, 
``Bogomol'nyi bounds and the supersymmetric Born-Infeld theory", 
 Nucl. Phys. B531 (1998) 168,  
 hep-th/9805054. }

\lr \abo{
A. Abouelsaood, C. Callan, C. Nappi and S. Yost, 
``Open strings in background gauge fields", 
 \np B280 (1987) 599.}
\lr \dlp{
J. Dai, R.G. Leigh and J. Polchinski,
``New connections between string theories", 
Mod. Phys. Lett. {A4} (1989) 2073.}

\lr \mera{ R.R. Metsaev  and  M.A. Rakhmanov,
``Fermionic terms in the open superstring effective action", 
    Phys. Lett. B193 (1987) 202.}

\lr \gs { M.B. Green and J.H. Schwarz,
``Covariant description of superstrings", 
Phys. Lett. B136 (1984) 367.}

\lr \gsh {  M.B. Green and J.H. Schwarz,
     ``Supersymmetrical dual string theory. 2.Vertices and trees",
Nucl. Phys. B198 (1982) 252;
J.H. Schwarz, ``Superstring theory", 
Phys. Rept. 89 (1982) 223.
 }

\lr\bag{ J. Bagger and A. Galperin, {``Matter couplings in
partially broken extended supersymmetry"},
\pl B336 (1994) 25, hep-th/9406217.}
\lr\bgg{ J. Bagger and A. Galperin,
{``A New Goldstone multiplet for partially broken supersymmetry"},
\pr D55 (1997) 1091, hep-th/9608177.}
\lref\pol{J. Polchinski,
{``Dirichlet Branes and Ramond-Ramond charges"},
\prl 75 (1995) 4724,
hep-th/9510017.}
\lr \cecf{S. Cecotti and S. Ferrara,
{``Supersymmetric Born-Infeld Lagrangians"},
\pl B187 (1987) 335. }

\lr\lei{R.G. Leigh, {
``Dirac-Born-Infeld action from the
Dirichlet sigma model"},
\mpl A4 (1989) 2767.}

\lr\jhs{M. Aganagic, C. Popescu and J.H. Schwarz,
{``D-brane actions with local kappa symmetry"},
Phys. Lett. B393 (1997) 311, hep-th/9610249;
{``Gauge invariant and gauge fixed D-brane actions"},
Nucl. Phys. B495 (1997) 99, hep-th/9612080.}
\lr\bert{E. Bergshoeff and P.K. Townsend, {``Super D-branes"},
Nucl. Phys. B490 (1997) 145,
hep-th/9611173.}

\lr\gib{ G.W. Gibbons and D.A. Rasheed,
{``Electric - magnetic duality rotations in nonlinear electrodynamics"},
Nucl. Phys. B454 (1995) 185,
hep-th/9506035.}

\lr\zumg{
M.K. Gaillard, B. Zumino,
{``Nonlinear electromagnetic selfduality and Legendre transformations"},
hep-th/9712103.}

\lr \zumm{D. Brace, B. Morariu  and  Bruno Zumino,
``Duality Invariant Born-Infeld Theory", 
hep-th/9905218, this volume. }

\lref \man{M. de Groot and P. Mansfield,
{``The Born-Infeld Lagrangian via string field theory"},
Phys. Lett. B231 (1989) 245. }

\lr \hup {J. Hughes and J. Polchinski,
{``Partially broken global supersymmetry and the superstring"},
\np B278 (1986) 147.}
\lr \hulp{
J. Hughes, J. Liu and J. Polchinski,
{``Supermembranes"},
\pl B180 (1986) 370.}

\lr \seth{S. Paban, S. Sethi  and  M. Stern,
``Supersymmetry and higher derivative terms in the effective action of Yang-Mills theories", JHEP 9806  (1998) 012,  
hep-th/9806028; 
``Constraints from extended supersymmetry in quantum mechanics", 
 Nucl. Phys. B534 (1998) 137,  
 hep-th/9805018. }

\lr \hashi{A. Hashimoto, ``The Shape of Branes Pulled by Strings",
 Phys. Rev. D57 (1998) 6441, hep-th/9711097.}

\lr \brec{D. Brecher, ``BPS States of the Non-Abelian Born-Infeld Action",
 Phys. Lett. B442 (1998) 117;
D. Brecher  and  M.J. Perry, 
``Bound States of D-Branes and the Non-Abelian Born-Infeld Action", 
 Nucl. Phys. B527 (1998) 121, hep-th/9801127.
}

\lr \gaun{J.P. Gauntlett, J. Gomis  and  P.K. Townsend,
``BPS Bounds for Worldvolume Branes", JHEP 9801 (1998) 003,
hep-th/9711205.}

\lr \fts{E.S. Fradkin and  A.A. Tseytlin, ``Effective action approach to                  superstring theory",                                                        
    Phys. Lett.  B160 (1985) 69.}

\lr\tse{ A.A. Tseytlin,
{``Selfduality of Born-Infeld action and Dirichlet three-brane
of type IIB superstring theory"},
\np B469 (1996) 51, hep-th/9602064.}

\lr \tsen { A.A. Tseytlin, {``On non-abelian generalization of
Born-Infeld action in string theory"},
Nucl. Phys. B501 (1997) 41, hep-th/9701125. }

\lr \tsep { A.A. Tseytlin, ``Black holes and exact solutions in string theory", 
 in  Proc. of the  International School of Astrophysics ``D. Chalonge": Current Topics in Astrofundamental Physics,
4-16 September 1994, Erice, ed. N. Sanchez   
(World Scientific, Singapore).  
     } 

\lr  \liut{H.  Liu  and  A.A. Tseytlin,
``Dilaton - fixed scalar correlators and $AdS_5\times  S^5$ - SYM correspondence", 
 hep-th/9906151. 
   } 

\lr \mrt{
 R.R. Metsaev, M.A. Rakhmanov  and  A.A. Tseytlin, ``The Born-Infeld
     action as the effective action in the open superstring theory'', 
    Phys. Lett. B193 (1987) 207.
 }

\lref \ft {E.S. Fradkin and A.A. Tseytlin, 
 ``Effective  field theory from quantized strings", 
 Phys. Lett. B158 (1985) 316;
``Quantum string theory                        
    effective action",                                                        
    Nucl.  Phys.  B261 (1985) 1.
}

\lr \andr{O.D. Andreev and A.A. Tseytlin, ``Partition function                             representation for the open superstring effective action:                 
    cancellation of Mobius infinities and derivative                          
    corrections to Born-Infeld lagrangian",                                     
    Nucl. Phys. B311 (1988)  205.    }    

\lr \andrt{O.D. Andreev and A.A. Tseytlin, ``Generating functional                              for scattering amplitudes  and effective action 
in the open superstring theory",                                      
     Phys. Lett.  B207 (1988)  157.    }

\lr \andre{O.D. Andreev and A.A. Tseytlin, 
 ``Two loop beta function in the                
    open string sigma model and equivalence with string                       
    effective equations of motion",                                             
    Mod. Phys. Lett.  A3 (1988) 1349.
}

\lr \fto{ E.S. Fradkin, A.A. Tseytlin, ``Fields as excitations of 
quantized coordinates", 
    JETP Lett. 41 (1985) 206 (Pisma Zh. Eksp. Teor. Fiz.
    41 (1985) 169).}

\lr \frts {E.S. Fradkin and A.A. Tseytlin,
{``Non-linear electrodynamics from quantized strings"},
\pl B163 (1985) 123.}

\lr \tsey{A.A. Tseytlin, ``Vector field effective action in the open                   superstring theory",                                                        
    Nucl. Phys. B276 (1986) 391;  Nucl. Phys. B291 (1987) 876 (E).}

\lr \dorn{H. Dorn, 
``The Mass term in non Abelian gauge field dynamics on matrix D-branes and T duality in the sigma
model approach",  JHEP 9804 (1998) 013,  hep-th/9804065; 
``NonAbelian gauge field dynamics on matrix D-branes",  
 Nucl. Phys. B494 (1997)  105, 
hep-th/9612120; 
 H. Dorn  and  H.J. Otto, 
``On T duality for open strings in general Abelian and non Abelian gauge field backgrounds", 
Phys. Lett. B381 (1996) 81,  
 hep-th/9603186.}

\lr \ts {A.A. Tseytlin, ``Renormalization of Mobius infinities and                   partition function representation for string theory                       
    effective action",                                                          
    Phys. Lett.  B202 (1988) 81.}

\lr \tseyh{ A.A. Tseytlin,
 ``On  SO(32) heterotic -- type I superstring duality in ten dimensions", 
 Phys.  Lett.   B367  (1996) 84. }

\lr \bik{ S. Bellucci, E.A. Ivanov and S.O. Krivonos, 
{``Partial breaking  $\N=4$ to $\N=2$: hypermultiplet as a Goldstone superfield"},
In: Proc. of  32nd Intern. Symposium on the Theory of Elementary 
Particles, 1--5 September, 1998, Buckow, Germany, 
hep-th/9809190.}

\lr\shmak{ M. Shmakova,  ``One loop corrections to the D3-brane action", 
 hep-th/9906239. 
  } 
\lr \zan{ A. De Giovanni, A. Santambrogio  and  D. Zanon,
``$\a'^4$ corrections to the  $\N=2$ supersymmetric Born-Infeld action", 
hep-th/9907214. } 

\lref\HK{A. Hashimoto and I.R. Klebanov, 
\pl B381 (1996) 437, hep-th/9604065.}

\lref\ceder{M. Cederwall, A. von Gussich, B.E.W. Nilsson and A. Westenberg,
``The Dirichlet super-three-brane in ten-dimensional type IIB supergravity'',
Nucl. Phys. {B}490 (1997) 163, hep-th/9610148; 
 M.~Cederwall, A.~von~Gussich, B.E.W.~Nilsson,
P.~Sundell, and A.~Westerberg,
``The Dirichlet super p-branes in ten-dimensional type IIA and IIB supergravity'',  Nucl. Phys. B490 (1997) 179,  hep-th/9611159.}

\baselineskip8pt
\Title{
\vbox
{\baselineskip 6pt{\hbox{   }}{\hbox
{Imperial/TP/98-99/67}}{\hbox{hep-th/9908105}} {\hbox{
  }}} }
{\vbox{\centerline {
Born-Infeld action, supersymmetry }  
\vskip4pt
 \centerline { and string theory    } 
}}
%
\vskip -32 true pt
\bigskip

\centerline{  A.A. Tseytlin\footnote{$^{\star}$}{\baselineskip8pt
e-mail address: tseytlin@ic.ac.uk}\footnote{$^{\dagger}$}
{\baselineskip8pt Also at  Lebedev  Physics
Institute, Moscow.}}

\smallskip
 \centerline {\it  Theoretical Physics Group, Blackett Laboratory,}
\smallskip
\centerline {\it  Imperial College,  London SW7 2BZ, U.K. }

\bigskip
\centerline {\bf Abstract}
\medskip
\medskip
\baselineskip12pt
\noindent
We review  and elaborate on 
some aspects  of Born-Infeld  action 
and its supersymmetric generalizations 
in connection
with string theory. 
Contents:
BI action  from   string theory; 
some properties of bosonic $D=4$  BI action;
\NO  and \NT  supersymmetric BI actions
 with manifest linear $D=4$ supersymmetry;
four-derivative terms in \NF supersymmetric  BI action;
BI actions with `deformed' supersymmetry  from  D-brane actions;
non-abelian generalization of  BI action;
derivative corrections to BI action in open superstring theory.

\bigskip
\vskip 60 true pt
\centerline{\it 
To appear in the Yuri Golfand memorial volume,  ed.  M. Shifman,
World Scientific  (2000) }
\Date {August 1999 }
\noblackbox
\baselineskip 16pt plus 2pt minus 2pt

\newsec{Introduction}
It is a pleasure for me  to contribute to Yuri Golfand's 
memorial volume.  I met  Yuri several  times 
during his occasional visits of Lebedev  Institute
in the 80's.  Two of our discussions in 1985 I remember
quite vividly.

 Golfand 
 found  appealing  the interpretation of string theory
as a theory of `quantized coordinates',  viewing it as a 
generalization of   some old ideas
of  noncommuting  coordinates. 
In  what should be an  early spring    
 of 1985 he read  our JETP Letter  \fto\ which was 
a brief Russian version of
 our  approach  with Fradkin  \ft\ to  string theory effective  action
based on  representation  
of   generating functional for string amplitudes 
as   Polyakov  string path integral with a 
covariant 2-d sigma model in the exponent.
In \fto\   our approach 
was  interpreted  in a somewhat  heuristic
way:
(i) the basic quantized `pre-field'   is a set of 
 coordinates $x^m$ of  a $D$-dimensional space 
which may be taken as a string (or membrane) coordinates
depending   on internal  parameters $\s^i$
(world-volume coordinates);
(ii) the classical space-time coordinates are expectation  or `center-of-mass' 
values of $x^m(\s)$; 
(iii) all space-time fields are `excitations'  of these 
quantized coordinates, appearing as generalized  `sources'  or `couplings'
 in the path integral action,  
 $\int d^2 \s \sum  \del^{k_1} x^{m_1} ... \del^{k_n} x^{m_n} B_{m_1 ...m_n} (x(\s))$. 
Golfand  asked me  if  superstring theory  should  then 
be interpreted as a theory  of quantized supercoordinates $(x,\theta)$. 
I told him of a recent paper by Green and Schwarz  \gs\ 
as the one that   should provide a basis for such  a program.
Later in  spring 1985  we generalized  the    sigma model 
approach  to Green-Schwarz superstring \fts\ 
using its light-cone  gauge formulation \gsh.

 In summer of  1985  Golfand approached  me again  after
having seen   the Lebedev Institute preprint version
 of  our paper \frts\
on the derivation of the  \BI action from the 
 open string theory. This 
 was a simple application of the
 non-perturbative in number of fields
approach of \ft, allowing one for the first time 
to sum certain terms 
in the string effective action to all orders in $\a'$.  
Yuri   stressed  the  importance of the  fact that  the 
 string tension $T=(2\pi \a')\inv$
is  determining the  critical value of  the electric  field.
He   was also excited about  a possibility (noted in \frts) 
of a kind of  `bootstrap'   if   such  a  non-linear 
action following itself from string theory 
admits  string-like solutions. 
In \frts\ we  mentioned that vortex solutions of 
similar \BI type actions (e.g.  $\sqrt{F^2}$)
 were discussed previously in \niel.
This idea   has  similarity to some recent 
  developments:
  a  simple   plane wave type  
solution of BI action ($F_{a0} + F_{a1}=0$)
may be re-interpreted  ($A_a = T X_a, \ F_{a1}=  T \del_a X_1$) 
as   a fundamental string ending on D-brane \refs{\calm,\gibb} -- 
the dimensional reduction of BI action 
along the 1-direction is simply the  DBI action \lei\  describing 
D-brane collective coordinates.\foot{This is of course not surprising 
from more stringy perspective,  as 
$T$-duality  along 1-direction  should  convert
the wave momentum into the (wound) string charge. Since this solution
is essentially a   plane wave,   it  
is also an exact solution of not only the  BI action
but also of the full  open string theory  effective action \thorl.}

I recall that Golfand was also  asking  me about  supersymmetric 
extension  of \BI action. At that time
I  was not aware of an early work  \des\ 
on this subject, but  later in 1986  there appeared 
the   paper  \cecf\  (inspired in part by the  discovery of the 
 relation of the bosonic  \BI action to  string theory)
  where  \NO  supersymmetric  version of $D=4$ 
BI  action was  presented in the explicit form. 
While the requirement of \NO $D=4$ supersymmetry did
not  fix uniquely 
the bosonic part of the nonlinear abelian vector multiplet action 
\cecf\ to be the standard 
$ \sqrt{ - \det (\eta_{mn} + T\inv F_{mn})}$, 
it  is now  clear  that  the condition of 
 $\N=4, D=4$ (or $\N=1, D=10$)
 supersymmetry should imply
 this.\foot{In particular, the structure of $F^4$ 
\refs{\bergr,\mera} and, more non-trivially, 
 $F^6$ \mrt\
 terms in the \NO supersymmetric 
deformation  of the $D=10$ super Maxwell action  was found 
to be exactly the same as in the BI action.} 

It seems, therefore, that a  review  of 
some aspects of the \BI action
  and  its
 supersymmetric extensions in the context of string theory
is    quite  appropriate  in this volume. 

\newsec{\BI action  from   string theory}

The \BI  action was   derived 
from  string theory in \frts\
as   field strength 
 derivative  independent part of the  open 
string effective action
by 
starting with an `off-shell'  bosonic open string path integral 
 on the disc
in an external (abelian) vector field 
\eqn\oct { 
 Z(A) = < \tr\ P \exp i\int d \vp \  \dot x^m A_m (x) > \ , }
$$     A_m(x_0 + \xi)  \to  \ha \xi^n F_{nm} + O(\del F) \ ,  $$
and using  specific ($\zeta$-function) renormalization
to get rid of a  linear divergence.
In Appendix we  repeat  the original computation \frts\ 
of this partition function in the abelian $F_{mn}=\const$ 
background  in a slightly generalized  form:  we shall assume that 
the   boundary part of the string action  contains  also  the 
usual  `particle' term  $M (\dot x^m)^2$.
Here the constant $M$ may be viewed   as an `off-shell' condensate
 of a massive  open string mode or simply as a 
formal regularization parameter. 
The effective boundary action will then 
have  both  a `first-derivative' scale-invariant 
($\sim T$)  and  a second-derivative ($\sim M$) 
 parts and  will  interpolate between the 
string-theory case $T\not=0, \ M=0 $ 
 discussed in \frts\  
  and the  standard particle  case   $T=0, \ M\not=0$
  appearing in the 
Schwinger computation of  $\log \det (-D^2(A))$. 
The resulting bosonic string partition function   for a single magnetic field component
$\f$   will be (superstring expression  is similar, see Appendix)
\eqn\ture{ Z \sim { \G ( 1+ { T  +  i\f \ov M} )\ 
\G ( 1 + { T  -  i\f \ov M})  \over
[ \G (1 + { T  \ov M}) ]^2 }   \ , 
 } 
 and thus will  have  the \BI 
$\sqrt{ 1 + (T\inv\f)^2}$ and  the 
Schwinger $ { \pi M\inv \f\ov {\sinh (\pi M\inv  \f)}}$ expressions
as its $M=0$ and $T=0$  limiting cases.

 The 
$F^2+\a'^2 F^4$ terms in BI action were  found to be in
precise agreement with the ones
derived  directly from (super)string 4-point amplitude
 \refs{\tsey,\gross}. The reason why the renormalized
  open string path integral on the disc in $F_{mn}=\const$ background
 reproduced, indeed, 
the correct effective action\foot{It could  seem  
that computing the  partition function we were not dividing 
over M\"obius group volume, compared to the standard 
on-shell generating functional for string S-matrix.}
was explained in detail 
later 
\refs{\ts,\andr}: 
the apparently missing M\"obius group volume factor
is only linearly divergent in the   bosonic 
case (and is  finite in the 
superstring case) 
 and thus is   effectively taken care of by the renormalization.
The  computation of the string partition function 
in  a constant  abelian  background is essentially an `on-shell'
computation: 
$F_{mn}=\const$ solves   equations 
of motion for any gauge-invariant 
 action $S(F)$ depending only  on the field strength
and its derivatives.\foot{Related point  is that the BI action is unambiguous:
it is not changed by local field redefinitions of gauge potential since 
these lead to  
   terms containing derivatives of $F_{mn}$
which, by definition, are not included in the BI action.
This is also related  
 to  the fact  that 
  the  string  partition function in 
$F_{mn}=\const$ background 
 does  not contain logarithmic  divergences.}

Furthermore, in the  important paper \abo\  it was  demonstrated
that the leading-order term  in the expansion in $\del F$ 
of  the  condition of conformal invariance of  the open string 
sigma model follows  indeed from the BI action.
In particular,  $F_{mn}=\const$  background  
defines  a  conformal 2-d  field theory.
The superstring generalization  of this  conformal invariance 
 argument 
implied \berg\
 that the derivative-independent  term in the open
 superstring effective action should  also  be given by 
the same BI action. After some initial  confusion  in \tsey\ 
(corrected in errata)
this conclusion  was   reached \refs{\mrt} also
  in the original  path integral approach of \ft.

\BI action  is a  unique example  of the  case 
when certain $\a'$ string corrections can be summed
to all orders. Though  no similar action is known 
in the closed-string context\foot{Apart from the suggestion in  \tseyh\ 
(based on type I -- heterotic string duality and  conjecture about 
special supersymmetry properties of the \BI action) 
that BI action may be summing up  $F^{2n+2}$   string 
 $n$-loop corrections 
in the heterotic string theory.}
the  expectation is 
that string tension defines a natural 
maximal scale  for all  field strengths, including  curvature.
The analogy 
with  open string theory
 suggests   that higher order $\a'$ terms in the effective action 
may eliminate 
 (at least   some)  black hole  singularities \tsep.
While in  the Maxwell 
theory the field of a point-like charge is singular at the origin and its energy is infinite, in the Born-Infeld theory  the electric 
field  of a $\d$-function source
is regular at $r=0$ (where it  takes its maximal value) 
and its total energy is finite \born. From the point of view of the distribution of the electric field ($\rho_{\rm eff}={1 \ov 4 \pi} $div $ E$) the source is no longer point-like  but  has an effective  radius  $r_0 \sim \sqrt {\a'}$ (for example, in 4 dimensions 
  $ \  E_r={F}_{rt}={
 Q\ov \sqrt{r^4 + r_0^4}} , \ \  r_0^2= 2\pi\a'  Q $).
Since both open and closed string  theories are effectively  non-local 
with  characteristic scale  of $\sqrt {\a'}$, 
one may expect that  Schwarzschild singularity is 
smeared in a similar way.

The remarkable
second  advent of D-branes \dlp\ 
 four  years ago \pol\  brought the \BI action  again into 
the  spot-light. 
In \lei\ the  \BI action  found a new interpretation  -- 
as  an action of D-branes    in the 
static gauge. What was  called `Dirac-Born-Infeld' action \lei\ 
was derived  by 
applying  the  conformal invariance 
conditions approach of \abo\ in the case 
of mixed (Dirichlet and Neumann) boundary conditions.\foot{Some subtleties
in the  approach of \lei\ and  attempts of 
generalization to the non-abelian case were discussed in \dorn.}  
The same action can be  also easily obtained \tse\ 
 using the path integral approach as in  \frts.\foot{Here the  
 aim  is to  compute
  the  string path integral on the disc 
 in the presence of a $D$-brane.
 This is the  partition function
of virtual open strings with    mixed Dirichlet-Neumann boundary 
conditions (i.e. with ends attached to a hyperplane)
 propagating in a  condensate  of massless  vector 
string modes. 
 The collective coordinates $X^i$ and internal vector 
$A_m$ degrees of freedom 
of the D-brane  are represented by the boundary  background 
couplings as in  \refs{\dlp,\lei}.}
The path integral approach makes 
T-duality covariance properties of the resulting D-brane actions 
transparent, implying that 
 all $p<  9$ brane actions can be  
obtained  by direct  dimensional reduction from the 
$D=10$ ($p=9$ brane)  \BI action. 
 Indeed, the  DBI action is  not a new action, but is simply the 
 reduction
of the BI action. In particular, all solutions 
of the DBI action can thus be obtained from the solutions 
of the higher-dimensional  BI action (see   \gibb). 

Thus the  form of   the 
D-brane  
action   
is  determined    
 by  the abelian $D=10$    open 
string effective action  
 \refs{\bac,\tse} and is given by 
 the \BI action for the $D=10$ vector potential\foot{Here
we use  the   Minkowski signature and  the following notation:
 $\mu,\nu=0,1,...,9$;\  $m,n= 0,1,...,p$; \ $s,r= p+1, ..., 9$, \ 
$T\inv = 2\pi \a'$, \ $\ep^{mncd}\ep_{mncd} =-4!$. 
 The functions $A_m$ and $X_s$ depend only on $x_n=(x_0,...,x_p)$.} 
$A_\mu  =(A_m, A_s= T X_s)$   reduced  down to $p+1$ dimensions\foot{In the low-energy or `non-relativistic' approximation, i.e. 
to the leading quadratic order in $F_{mn}$,  this action
  is the same as the 
dimensional reduction  of the $D=10$  $U(1)$ 
Maxwell action for $A_m$  \witt.
A simple determinant identity shows that this is true in general
for the whole BI action.}
$$S_p = \T_p \int d^{p+1} x \sqrt {-\det 
(\eta_{\m\n}  +   T\inv F_{\m\n})}  $$ 
\eqn\qer{
= 
 \T_p \int d^{p+1} 
  x \sqrt {-\det 
(\eta_{mn}  + \del_m X^s \del_n X^s +   T\inv F_{mn})}\  .  }
This  `T-duality' relation
suggests  that 
 supersymmetrization of the DBI action  (and its non-abelian generalization)
in flat space\foot{This relation may 
 no longer apply  in  the case  of  a non-trivial 
closed string background  without  simple  isometry properties.} 
 should  
also be determined  by that of the \BI action.

Originating from the BI action, the DBI action 
implies similar `maximal field strength'  constraints on allowed
physical configurations. 
In particular, the action for D0-brane is simply that of a relativistic 
particle $\int dx_0 \sqrt{ 1 - (\del_0 X^s)^2}$, and 
the ` maximal field strength' constraint  here is simply the standard 
relativistic  constraint on particle's   velocity \bac. 
Here it is interesting to recall  
 that  it was the analogy with the  square root structure
of the 
relativistic particle  action that  was 
one of the original motivations  of Born in looking for 
a non-linear electrodynamics action \born\ 
which  does not allow the electric field of a point charge 
to  become infinite.

\newsec{Some properties of bosonic $D=4$  \BI action}
The $D=4$  \BI    Lagrangian 
\eqn\bia{
L_{\rm BI}=  \sqrt{- {\rm det}_4  (\eta_{mn} + F_{mn})} - 1 \ , }
where we set the fundamental (scale)$^2$(=$T\inv = 2 \pi\a'$ in the string theory context)
 equal to 1, 
has several remarkable features, including 
electro-magnetic  duality  
 and causal propagation (see, e.g.,  \refs{\gib,\deser}
 and refs. there). 
Since in four  dimensions
\eqn\foo{ -{\rm det}_4 (\eta_{mn} + F_{mn}) = 1 + \ha F_{mn}F^{mn} -
{\textstyle {1\ov 16} } (F_{mn}\sF^{mn})^2 \ , \ \ \ \ \
\sF^{mn}\equiv \ha\ep^{mncd} F_{cd} \ , }
$L_{\rm BI}$  interpolates between  the Maxwell Lagrangian 
$\four F_{mn}F^{mn}$  for small $F$  and the total derivative (topological
density) 
${i\ov 4}  F_{mn}\sF^{mn}$ for  large $F$.
In Euclidean signature  one finds  \gibb\ 
\eqn\euc{
L_{\rm BI}=  \sqrt{(1 + \four F_{mn}\sF^{mn})^2 + \four (F_{mn} -\sF_{mn})^2}
  -1\   \geq \  \four F_{mn}\sF^{mn} \ , }
implying that the minimum of the Euclidean  action 
is attained  at  (abelian) self-dual  fields
(for a discussion of related  BPS bounds for  DBI actions 
see \gaun). 
Another useful representation is 
\eqn\euci{
L_{\rm BI}=  \sqrt{(1 + I_2)^2 +  2 I_4 } -1  
= I_2 + I_4 [1 + O(F^2) ] 
  \ , }
$$ I_2 \equiv  \four F_{mn}F^{mn} \ , \ \ \ \ \  
I_4 \equiv 
 - \eight\bigg[ F_{mn} F^{nk} F_{kl} F^{lm} - \four (F_{mn}F^{mn})^2 \bigg] 
 = - \eight (F^{(+)})^2 (F^{(-)})^2 \ . $$
That $L_{\rm BI} = I_2 + I_4 + O(F^6)$ 
is true  in all dimensions, but it is only in $D=4$ that all 
higher order terms are proportional 
to $I_4$.  This fact is reflected in the structure of the 
supersymmetric generalization of the $D=4$ BI action
(see section 4.1).

The  $D=4$ \BI action is obviously symmetric under 
$F \leftrightarrow F^*$ and is also covariant under the electric-magnetic 
(or vector $\to$ vector) duality,
 as can be concluded  from the structure of the equations of motion 
\gib\ (see also  \refs{\zumg,\zumm})
or demonstrated directly at the level of the action by 
following the standard steps of introducing 
 the 
Lagrange multiplier for the $F= dA$ constraint 
and solving for $F$ in the classical approximation 
\tse.

In four dimensions  it is possible  to write down 
the BI action in the form quadratic in $F_{mn}$ 
 by introducing two complex  auxiliary scalar fields \rot.
First, we replace $L_{\rm BI}$ (changing its overall 
sign as appropriate for the  Minkowski signature) by
$$ - \sqrt{-\det (\eta_{mn} + F_{mn})} \
\to \ \ -\ha V \det (\eta_{mn} + F_{mn}) + \ha V\inv\ ,  $$
  use \foo\  and 
introduce the second auxiliary field $U$ to `split'
 the quartic $(FF^*)^2$ term .
Finally, we can eliminate the term with $V^{-1}$ by
introducing a complex auxiliary scalar $a=a_1+ia_2, \
\bar a=a_1-ia_2$, and writing the Lagrangian as \rot
\eqn\iii{ L_{4} = -\ha V (a + \bar a + \bar a a -
\ha F_{mn}F^{mn} ) + \ha U [i(a-\bar a)
+ \ha F_{mn}\sF^{mn}] - \ha (a + \bar a ) \ , } or as 
\eqn\ciii{ L_{4} = -  {\rm Im} \bigg(\l \bigg[ a +
\ha \bar a a - \four (F_{mn}F^{mn} +
i F_{mn} \sF^{mn})\bigg] + i a\ \bigg) \ , }
$$
\l = \l_1 + i\l_2 \equiv U + i V \ . $$
The constraint implied by $\l$ is solved by $a=a(F)$ with
Im $  a (F) = \four F^{mn} F^*_{mn}$ and 
the real part
\eqn\solv{
{\rm Re} \ a(F)\ = \ 
\sqrt { 1 + \ha F^2 - {\textstyle{1 \ov 16}} (F\sF)^2 } -1 \ , }
which (up to sign) is    the
BI Lagrangian itself. This gives a natural `explanation' for the square
root structure of the \BI action. One can thus view the $D=4$  BI action as
resulting from a peculiar theory  for two  complex non-propagating
scalars $(\l,a)$ coupled non-minimally to a vector.
Shifting $\l$ by $i$
the Lagrangian \ciii\
can be put into the  form  
that does not contain terms linear in the fields
\eqn\cjj{
L_{4} = - \four F_{mn}F^{mn} + \ha \bar a a - {\rm Im}\bigg( \l \big[ a +
\ha \bar a a - \four ( F_{mn}F^{mn}
+ i F_{mn} \sF^{mn})\big] \bigg) \ . }
Since in this form the BI action is quadratic
in the vector field, it is very simple to demonstrate its covariance
under the vector-vector  duality.
Adding the Lagrange multiplier term $ \ha \td F^{*ab} F_{ab},$
where $\td F_{ab}$ is the strength of the dual vector field, 
and integrating out $F_{ab}$ we find that the dual action has the same
form as \ciii\ with \foot{The equations of motion derived
from the
vector terms in the action \ciii\ have the full
$SL(2,R)$ invariance:
$
\l \to { p \l + q \ov k \l + l } , $ $ \ \
F_{mn} \to (k U + l) F_{mn} + k V F^*_{mn} , $ $\ pl-qk=1,
$ see also \refs{\zumg,\zumm}.}
\eqn\dual{F_{mn} \to \tF_{mn} \ , \ \ \ \ \ \ \
\l \to - { \l\inv }\ , \ \ \ \ \ \ \ \ \ a \to - i \l a \ . }
Like  the Maxwell  action, the action \ciii\ is not invariant under this
duality. There exists, however, an equivalent
action containing one extra vector field variable
which is manifestly duality-symmetric \rot.
                                                                                                                                        duality-symmetric actions was explained

The BI  Lagrangian  in the form \ciii,\cjj\ 
 may be viewed as a special case of the following
Lagrangian  for a vector coupled non-minimally to  a set of massive scalars
\eqn\acc{
L = - \four F_{mn}F^{mn} -\ha (\del_a \vp_i)^2 - 
\ha m^2_i \vp_i^2 +
g_{ijk} \vp_i \vp_j \vp_k +
\vp_i (\a_i F_{mn}F^{mn} + \b_i F_{mn} \sF^{mn}\big) \ . }
In the limit when 
masses of scalars are much larger than their gradients
so that the $(\del_a \vp_i)^2 $ terms may be ignored,
\acc\ reduces to \cjj\  with
the scalars $\vp_n$ being linear combinations of $\l_1,\l_2,a_1,a_2$
in  \cjj. 
This action may be viewed as a truncation of the cubic open string field
theory action
which reproduces the BI action as an effective action
upon integrating out at the string tree level all 
massive string modes (represented here by $\vp_i$)
 \refs{\frts,\man}. The kinetic
terms  $(\del_a \vp_i)^2 $
may be dropped since they  lead to  derivative-dependent
$O(\del F )$ terms which, by
definition, are not included in the
leading part of the low-energy effective  action.
Note that to  represent  higher dimensional \BI  action 
in a cubic form similar to \acc\ one would need to introduce
auxiliary tensor fields to `split' the 
higher-order $F^k$ invariants in $\det(\eta_{mn} + F_{mn})$. 

\newsec{Supersymmetric \BI actions
 with manifest  $D=4$ supersymmetry}
Below we shall review what is known about 
generalizations  of  \BI action 
with  manifest $D=4$  supersymmetry. 

There exists a remarkable connection between \ \ 
(i) partial supersymmetry breaking, \ (ii) nonlinear realizations of
extended supersymmetry, (iii) BPS solitons, and (iv)
nonlinear Born-Infeld type actions  (see, e.g.,  \refs{\hup,\hulp,\bgg,\rot}
and refs. there). Extended 
$\N> 1 $ supersymmetry can be partially
broken either by a translationally non-invariant
background (soliton) in a second-derivative higher-dimensional theory
or by a translationally invariant vacuum
in a nonrenormalizable
theory in four dimensions containing non-minimal interactions \apt.

The interpretation (and derivation) of  $\N=1$ supersymmetric  BI action
\refs{\des,\cecf}  as the action for a 
Goldstone multiplet  associated with partial breaking
of $\N=2 $ to  $ \N=1$ supersymmetry was suggested  in \bgg.
As was demonstrated   in \rot,  the connection between
partial breaking of supersymmetry and nonlinear actions
is not accidental and has to do with constraints that
lead directly to nonlinear actions of \BI type.
Spontaneously broken symmetries give nonlinear realizations of the
broken symmetry group. A standard  way to find such realizations is
to begin with a linear representation and impose a nonlinear
constraint.
The  constrained superfield approach \refs{\martin,\rot} 
appears to be  a  universal
and  transparent way of deriving and dealing with   these actions.

In a similar way,  a 
 massless  $\N=2$  vector multiplet may  be  also considered as a 
{Goldstone} multiplet associated with  partial spontaneous
breaking  of $\N=4 $  supersymmetry
to  $ \N=2$   \bik.  The  $\N=2$ analog of the   $\N=1$ supersymmetric \BI 
action  was suggested in \ket.  Though this was not proved to all orders, 
it  is likely that 
the  bosonic part of this action 
  is related (after  a field  redefinition
eliminating   higher derivative scalar terms) 
to the DBI action for a 
3-brane  moving in 6 dimensional space-time 
 (with two scalars of the  $\N=2$ vector multiplet
playing the role of the transverse collective coordinates).

The $\N=4$ supersymmetric extension of the \BI action 
(written, e.g.,  in terms of $\N=1$ or $\N=2$  superfields)
 is not known at present 
and it appears to be non-trivial to  construct it (cf. \rot). 
Below we shall describe  what can be  learned  
about the structure of the $F^4$ term in such $\N=4$ action 
using the knowledge of the $\N=1$ and $\N=2$  \BI actions 
and assuming the expected global $SU(3)$ symmetry 
of the $\N=4$ action written in terms of $\N=1$ superfields \liut. 
 
\subsec{ \NO supersymmetric  action}
The  $\N=1$ supersymmetric BI action 
can be written 
 in the following way \refs{\cecf}\foot{We follow the same 
conventions as in 
\rot, in particular, 
$D^2=\ha D^\a D_\a$.} 
\eqn\solt{S  = \ha \int d^4 x ( \int d^2 \t\  W^\a W_\a + h.c.) 
+  \int d^4 x \int  d^2 \t d^2 \bar \t  \
 B(K, \bK) W^\a W_\a \bar W^\da \bar W_\da  \ , } 
where
\eqn\geg{ B\equiv { 1 \ov 1 - \ha (K + \bK)   + \sqrt { 1 - (K + \bK) 
 + \four (K - \bK)^2 } } = \ha   + \four (K + \bar K) + ...   \ , }
\eqn\deeq{K\equiv D^2 (W^\a W_\a ) \ , \ \ \ \ \ \
\bK\equiv  \bD^2 (\bar W^\da \bar W_\da )\ .   }
The action depends in general on dimensional scale parameter which is set to 1 as in \bia. 
Since 
 $$(D^2 W^2)_{\t,\bt=0} = - \four
F^{ mn}F_{ mn} - { i \ov 4} F^{ mn}\sF_{ mn} -
 {\te {i\ov 2}}  \l \s^m\del_m \bar \l  + \ha {\rm D}^2 $$
the bosonic
part of the
action depends on the square of 
 the auxiliary field $ {\rm D}$  so that
${\rm D}=0$ is always a solution.

To get insight into the structure of \solt\ 
and to exhibit its invariance under  the
 second (spontaneously broken, i.e. 
 non-linearly realized)  supersymmetry \bgg\ 
it is useful to rederive this non-linear action 
using constrained  $\N=2$ superfield approach \rot.
We start with  $\N=2$ vector multiplet  described by  constrained chiral field
strength $  \W(x,\t_1,\t_2)$ that obeys the Bianchi identity 
$D^2_{ab}  \W=C_{ac}C_{bd}\bar D^{2cd}\bar   \W$ \ ($a,b=1,2$).
Defining  $D\equiv D_1 ,\ Q\equiv D_2$ this becomes
$D^2  \W=\bar Q^2\bar  \W ,\  DQ  \W=-\bar D\bar Q \bar  \W.$
We break $\N=2$ supersymmetry to $\N=1$ by assuming that $  \W$ has a
Lorentz-invariant condensate $\exw$ (we set the scale of the
 supersymmetry breaking to $1$):
\eqn\ntwocon{  \W \to  \exw + \W \ , \ \ \ \ 
\exw= -\t^2_2\ ,\ \ \ \ \ \langle Q^2  \W \rangle = 1\ ,\ \ \ \ \
D\exw= 0\ . }
We reduce the
field content to a single $\N=1$ superfield by imposing\foot{This removes 
  independent  chiral superfield part of the $\N=2$ multiplet.
It would be interesting  to generalize the discussion to the case when 
the chiral superfield  is  first 
kept independent and then  integrated  out.}
\eqn\nww{\W^2=0\ .} 
Then the  above constraints    imply 
\eqn\nltr{Q^2\W =\bar D^2 \bar\W - 1\ ,
\ \ \ \ \ \ \  0=\ha Q^2\W^2=\W(\bar D^2 \bar\W - 1)+
\ha Q^\a\W Q_\a\W\ .  }
Projecting to $\N=1$ superspace by setting $\t_2=0$ and defining the
$\N=1$ superfields
\eqn\nldef{  \P\equiv\W|_{\t_2=0}\ ,\ \ \ \ \ \ \
W_\a\equiv-Q_\a\W|_{\t_2=0}\ ,}
we find that the  constraint
for the chiral superpartner of the vector multiplet in the $\N=1$
superspace description of the $\N=2$ vector multiplet is 
\eqn\vecon{  \P=\P\bar D^2{ \bar \P}+\ha W^\a W_\a\ , }
which  coincides  with  
 the constraint in  \bgg. 

Because of the constraints \ntwocon,\nww,\vecon, there are many
equivalent   $\N=2$ forms that all give the same  $\N=1$ 
action \rot. One example is  a 
class of actions  proportional to the $\N=2$ Fayet-Iliopoulos term:
$\int d^2 \t_1 d^2 \t_2\ \cF (  \W)=  \cF'' (0) \int d^2 \t_1  \P$. 
Another action that leads to  the same constraints and the  final $\N=1$
Born-Infeld action is the standard free $\N=2$ vector action,
i.e. the  action for the $\N=1$   vector ($V$) and chiral ($ \P$)
superfields, plus a  term with a chiral $\N=1$
superfield Lagrange multiplier $\L$ imposing the constraint \vecon\  
\eqn\iui{ S= \int d^4 x \bigg( \int d^2 \t \bigg[ ( \ha W^\a W_\a +  \P \bD^2
{ { \bar \P}} ) + i \L (\ha W^\a W_\a +  \P \bD^2 { { \bar \P}} -  \P) \bigg] + h.c. \bigg) \ . }
Shifting $\L \to \L + i$, we get 
\eqn\tyt{ S = \int d^4 x \bigg[ \int d^2 \t \bigg(
i\L \big[\ha W^\a W_\a +  \P \bD^2 { { \bar \P}} -  \P\big] \ +\  \P\ \bigg)
+\ h.c. \bigg] \ . }
The resulting action  is thus simply
\eqn \pti{ S= \int d^4 x \bigg[ \int d^2 \t \  \P(W,\bar W) +
h.c. \bigg] \ ,  }
where  $ \P$ is the solution of the constraint \vecon. 
Since the  explicit solution of \vecon\ is \bgg
\eqn\solu{ \P (W,\bar W) = \ha W^\a W_\a
+\ha \bD^2 \bigg[ B(K, \bK) W^\a W_\a \bar W^\da \bar W_\da\bigg] \ , } 
where  $B$ was defined in \geg,
 the action \pti\  is 
nothing but the  $\N=1$ supersymmetric \BI action \solt.

One concludes \refs{\bgg,\rot} that 
 the requirement of partially broken $\N=2$ supersymmetry
uniquely fixes the action for the   $\N=1$ vector multiplet to be the
supersymmetric Born-Infeld action.
As  explained in \rot,   the $\N=1$ supersymmetric  Born-Infeld
action  also emerges as an effective action 
from the  $\N=2\to \N=1$
supersymmetry breaking  model of \apt\ when one decouples (`integrates
out') the massive chiral multiplet.

The bonus of the  above derivation 
is that it reveals the  hidden non-linearly realized supersymmetry
of the \BI action \pti\ -- the broken 
 half of the original $\N=2$ symmetry 
\refs{\bgg}. 
The second ($\N=2$)  supersymmetry transformation law
follows from the  above constraints and definitions of the $\N=1$ superfield
components, 
\eqn\trans{\d_2 \P\equiv(\eta^\a Q_\a+\bar\eta^\da \bar Q_\da)\W|_{\t_2=0}
=-\eta^\a W_\a\ ,\ \ \ \ \ \d_2W_\a=\eta_\a(\bar D^2\bar \P - 1)-i\bar\eta^\da
\partial_{\a\da}\P\ .}
Note that this transformation is non-linear since 
 $\P$  \solu\ contains terms of all orders in $W$
(and thus also in the fundamental scale parameter or in $2\pi\a'$). 

 The bosonic part of the supersymmetric action \tyt\ is
exactly the BI action represented in the form \ciii\ with the 
two auxiliary
complex scalar fields  fields $a$ and $\l$
being the corresponding scalar
components of the chiral superfields $\P$ and $\L$ in \tyt.
As in the bosonic case \ciii,
the  Lagrange multiplier representation \tyt\ of the action 
also simplifies \rot\  the proof \bgg\ of the {duality covariance}
of
the $\N=1$  supersymmetric 
\BI action \pti,\solu.

Let us make  a brief  comment about  the quantum 
properties of the \BI actions. Viewing BI action as 
a leading term in the low-energy
effective action of string theory,   it does not  make 
much sense to  quantize it directly,\foot{Ignoring 
derivative corrections, 
\BI action  represents a  sum of string tree diagrams with massive 
modes on internal lines and massless vectors on external lines. 
Quantum loop corrections to \BI action thus represent only a subclass
of all string loop diagrams where, e.g., loops of massive modes 
are not included. It is only the sum of all string diagrams 
at a given loop order that is expected to be UV finite.}
unless one is systematically keeping all momenta  small 
compared to the cutoff $1\ov \sqrt{\a'}$  as 
in other effective field theories (see, e.g., \don).    
 Still, formally, one may try to view the BI action as defining 
a fundamental theory and 
 compute the corresponding 
 quantum corrections   using, e.g., 
background field method. It is easy to see that {\it logarithmically} 
divergent
corrections 
to the abelian bosonic BI action  will involve derivatives 
of the field strength, i.e. the original  BI action is not renormalizable.
The same is true in the  $\N=1$ supersymmetric case:
as follows  from \solt,  all terms  additional  to the Maxwell $W^2$  term 
in the $\N=1$ BI Lagrangian are 
no longer F-terms, but  D-terms, and thus 
may  be deformed by quantum corrections.\foot{Note, however, 
that there should be no finite quantum renormalization 
of the coefficient in front of the $\N=1$ BI  action  \solt\   as 
part of the full quantum effective action:
assuming that the second spontaneously broken  supersymmetry 
survives at the quantum level, it should  again relate the 
coefficient of the D-term in \solt\
to that of the   F-term  and thus  should rule
 out its  finite renormalization.
This should be related to expected non-renormalization
of the BPS  3-brane tension.}
This was  indeed confirmed by explicit computations in \refs{\shmak,\zan}
which demonstrated 
that the leading  1-loop logarithmically divergent 
correction to the   ($\N\geq 2$) supersymmetric \BI action 
has the  $\del^4 F^4$ form.
 It is not  completely  surprising that  the  same  4-derivative term 
 (whose $\N=1$ structure $\sim \int d^4 \t \del_m W^\a \del^m W_\a \del_n \bar W^\da \del^n \bar W_\da$ \zan\ 
  is 
 determined by the supersymmetry) 
appears  as the leading derivative correction to the \BI term  
in the tree-level open superstring effective action \andr\ (see section 7).

\subsec{ \NT supersymmetric  action    }

The  \NT extension of the  \BI action  suggested in \ket\ 
is  similar in structure to the  \NO one \solt:
\eqn\aact{
 S=  \ha  \int d^4 x \bigg[   ( \int d^4 \t \  \W^2 + c.c.) 
+\four \int d^4 \t  d^4 \bt  \  {\cal B}(\K,\bar \K) 
 \W^2 \bar \W^2 \bigg]  \ , }
where
 \eqn\uiut{ {\cal B}= { 1 \ov 1  - \ha (\K + \bbK)   + \sqrt { 1 -
 (\K + \bbK) 
 + \four (\K - \bbK)^2 } }  \ , }
\eqn\diq{\K\equiv \ha D^4 \W^2 \ , \ \ \ \ \ \
\bbK\equiv \ha  \bD^4 \bar\W^2\ . }
The Lagrangian here $H= \W^2+...$ (cf. \pti) 
satisfies  the \NT generalization of the \NO 
non-linear constraint  \vecon, 
\eqn\coon{ H = \four H \bar D^4\bar  H  + \W^2 \ ,
 \ \ \ \ \  \ \ \ \   H^2=0 \ .  } 
The analogy with the $\N=2 \to \N=1$ supersymmetry  
breaking case  discussed above
 suggests  a relation to  the  $\N=4 \to \N=2$ supersymmetry 
breaking  \refs{\bik}
 and thus the  interpretation of \aact\ 
 as the unique action  for the  $\N=2$ vector
multiplet as a Goldstone multiplet associated with 
the partial breaking of  $\N=4$ supersymmetry.
In this case \aact\ should have 
 hidden invariance
 under  two extra  spontaneously broken and non-linearly 
realized supersymmetries which should 
unambiguously determine the  form of the 
 action via the non-linear constraint \coon.

This  action  contains terms without derivatives and with 
 higher derivatives of the complex  scalar field. At first sight, 
this  seems 
to contradict its possible interpretation as a  DBI action 
(the Nambu-type  actions  for the transverse 
collective 
coordinates should contain scalars  only through their
first  derivatives   as required by  the  translational invariance).
However, it is likely that the higher-derivative terms 
can be eliminated  by field redefinitions (cf. \refs{\ket,\roch,\rot}). 
These, however, 
will make $\N=2$ supersymmetry of the
 resulting action non-manifest.  This
clash between the requirement of dependence on 
 first derivatives of scalars
  and manifest  extended  supersymmetry 
 is likely to be the general property 
of the $\N \geq 2$ supersymmetric \BI actions.

To  demonstrate that such unwelcome  terms 
 can indeed be redefined away
  let us consider 
the first subleading  four-field term  in the  action  \aact\
and show that  the second-derivative scalar terms there are indeed 
proportional to the leading-order equation of motion $\del^2 \p$
 (i.e. vanish on shell) and thus can be eliminated 
by  a field redefinition. 
The  $\N=2$ chiral superfield $\W$
satisfies the   constraints 
$
\bD_{\ad i} \W =0  , \ 
D^4 \W = \del^2 \bar{\W}  ,  
$
implying  the following expansion in terms of $\N=1$ superfields:
$$ 
\W =  \, \P (\td{y}, \t) + \sqrt{2}   \t_2^\a W_{\a}(\td{y}, \t) + 
  \t_2^\a   \t_{2\a} \bD^2 \bar{\P} (\td{y}, \t) 
 $$
\eqn\supfex{
=  \, \P(y, \t) + i   \t_2 \s^{m} \bar \t_2 \del_{m} \P(y, \t)
+ \four   \t_2^2 \bar \t_2^2 \del^2 \P +
\sqrt{2}   \t_2^\a W_{\a}(y, \t) -{i \ov \sqrt{2}}   \t_2^2 
\del_{m} W \s^{m} \bar \t_2 +   \t_2^2 \bD^2 \bar{\P}(y, \t)
\ , }
where $\t\equiv \t_1$  and  $\td{y} = y + i   \t_2 \s \bar \t_2
= x +  i \t \s \bar{\t} +  i   \t_2 \s \bar \t_2 .$
 $\P$ has  the  standard $\N=1$ component expansion
\eqn\eeeq{
\P =\vp  + i \t \s^{m} \bar{\t} \del_{m}\vp 
+ \four \t^2 \bar{\t}^2 \del^2\vp +
\sqrt{2} \t \psi  - {i \ov \sqrt{2}} \t^2 
\del_{m} \psi \s^{m} \bar{\t} + \t^2 {\cal F}
\ . } 

The  leading correction term in \aact\ is 
\eqn\actex{  
{\II}_4  = \int \! d^4 \!   \t \  {\bf I}_4\ , \ 
\ \ \ \ \ \ \ \     {\bf I}_4  =  \int \! d^4 \!   \t_2 \  \W^2 \bar{\W}^2 \ , 
} 
where ${\bf I}_4 $ can be expressed  in terms of $\N=1$ fields as follows
\eqn\supeno{\eqalign{
 {\bf I}_4 & =W^{\a} W_{\a} \bar{W}_{\ad}  \bar{W}^{\ad} 
+ i (\bP \del_{m} \P - \P \del_{m} \bP) W^{\a} \s^{m}_{\a \ad} 
\bW^{\ad}
\cr
& + 
\ha (\bP \bP \del_{m} \P \del_{m} \P + \P \P \del_{m} \bP \del_{m} \bP)
-2 \P \bP \del_{m} \P \del_{m} \bP \ .  }}
The cross-term can be written  also as 
$W^{\a} \bW_{\ad} D_\a \P\bar  D^{\ad } \bP$. Related expressions for the $\N=2$ invariant $\W^2 \bar \W^2$   in terms of $\N=1$ superfields appeared in \refs{\roch,\ket}.
If one is 
allowed to integrate by parts (which is possible  in the action 
under   the integral over $x$-space)  and 
omit terms proportional to equations of motion
(which can be redefined away by a transformation preserving $\N=1$ supersymmetry as in \roch)
then the  $\P^4$ terms 
in \supeno\ reduce to  just one term only, since
\eqn\alta{- \del_{m} ( \P^2 ) \del_{m} (\bP^2 )
+ \four \del^\m \del_\m ( \P^2 \bP^2) 
= 2 \del_{m} ( \P \bP ) \del_{m} (\P\bP )
- { 3 \ov 4} \del^\m \del_\m ( \P^2 \bP^2)  \ . } 
Integrating over $\t$  one finds that 
the component form of  the $\P^4$ terms  in \supeno\
agrees with  the 4-derivative term
 $(\del \vp )^2 (\del \bar \vp)^2$
in  the non-linear action  for a chiral multiplet in \refs{\bag,\rot}.\foot{Note that 
 the expansion of the full bosonic action is  
 $$ 
L = 1  + \ha \del_m \vp \del_m \bvp   +\four  F^2 
  -
 \eight [ F^4 - \four (F^2)^2] 
 - \ha (  F^2_{mn}  - \four F^2 \d_{mn} ) \del_m \vp \del_n \bvp
-   \eight \del_{m} \vp \del_{m} \vp
\del_{n}\bvp \del_{n}\bvp
 ,   $$
where,   modulo a  total derivative, 
$\del_{m} \vp \del_{m} \vp
\del_{n}\bvp \del_{n}\bvp  = 2 (\del_{m} \vp \del_{m} \bvp)^2   $.}

To all orders the scalar part of the $\N=2$ action 
is expected to coincide (after field redefinitions) 
 with 
$$L=\sqrt{\det (\d_{mn} + \del_m X^1 \del_n X^1 + \del_m X^2 \del_n X^2)} $$
which can be written in the form ($\vp \equiv X^1 + iX^2$) \rot\ 
$$L(\vp) =\sqrt{ 1 + \del \vp \del {\bar \vp}
+ \four (\del \vp \del {\bar \vp})^2
- \four (\del \vp)^2 (\del {\bar \vp})^2 } $$
\eqn\iuy{=  1 +  \ha \del \vp \del {\bar \vp}
-  { \four (\del \vp)^2 (\del {\bar \vp})^2 \ov
1 + \ha \del \vp \del {\bar \vp} +
\sqrt{ (1 + \ha \del \vp \del {\bar \vp})^2
- \four (\del \vp)^2 (\del {\bar \vp})^2 } } \ .}
This is the  bosonic part of the action of the non-linear chiral multiplet 
\refs{\rot,\bggg,\roche} (dual to a tensor multiplet action 
\refs{\bggg,\rot,\roche}) with $\N=1$ superfield Lagrangian 
\eqn\acvp{
L(\P) =\P\bar\P+{\ha (D^\a \P D_\a \P)( \bD^\da \bar\P\bD_\da \bar\P) \ov
1 + A 
+\sqrt { (1 + A)^2 -B  }}\ , }
$$A\equiv \del^\ada\P\del_\ada\bar\P\ , \ \ \ \ \ \ \ \ \ \
B\equiv (\del^\ada\P\del_\ada\P)
(\del^\ada\bar\P\del_\ada\bar\P)\  . $$
The Lagrangian  \acvp\ has
manifest {translational symmetry}  and thus defines   the action
representing  the 3-brane in 6 dimensions of ref.  
\hulp. It is natural to  expect that there exists
 an  exact  $\N=1$ superfield 
redefinition that puts the $\P$-dependent part of the $\N=2$ BI action \aact\
into the form \acvp. 

\subsec{ Four-derivative terms in the \NF supersymmetric  \BI action }
 It would be interesting for several reasons
(e.g., in connection with quantum properties of D3-branes
and their comparison with supergravity)
 to find a manifestly supersymmetric formulation
of $\N=4$ \BI  action generalizing $\N=4$ Maxwell theory. 
Related (by a field redefinition) component  action  
with $4$ linearly realized and $4$ nonlinearly
realized global $D=4$ supersymmetries can be  found 
by fixing static gauge and   $\k$ symmetry gauge 
in the D3-brane action
with global $D=10$ supersymmetry constructed 
in
\refs{\ceder,\jhs,\bert} (see section 5).
However,   the 
form of the $\N=4$ supersymmetric  action
with manifest  undeformed linear 
 $D=4$ supersymmetry, e.g., written 
 in terms of unconstrained $\N=1, D=4$ 
superfields -- one real vector with field strength $W_\a$ 
and 3 chiral scalar $\P_a$ ($a=1,2,3$),  is not known.  
For  $\P_2,\P_3$ set equal to zero the action  should reduce 
to the $\N=1$  form of the $\N=2$ action \aact. 
After a field redefinition eliminating higher derivative scalar 
terms the bosonic part of the action  should become  the 
$10\to 4$ dimensional reduction of the $D=10$  \BI action, i.e. the  
DBI action of a  D3-brane moving in  10 dimensions, 
\eqn\one{ S=  \int d^4 x \sqrt{- \det( \eta_{mn}
+ \del_m X^s \del_n X^s + F_{mn})} \ , \ \ \ \ \ \ \ \ \ s=1,...,6 \ ,  }
or, equivalently, squaring  the matrix  which 
appears under
 the determinant, 
\eqn\dbi{
S= \int d^4 x  \bigg[- \det ( \eta_{mn}  + 2 t_{mn} + t_{mr} t_{nr}
   +  F_{mr}F_{nr}  +2  t_{r (m} F_{n) r})\bigg]^{1/4} 
\ ,\  }
where $  t_{mn} \equiv   \del_m X^s \del_n X^s $.

The 6 real coordinates  $X^s$
should be related to the 3  complex  scalar components of $\P_a$ 
by 
\eqn\vov{
\vp^a = X^a + i X^{a+3}\  ,  \ \ 
\ \ \ \ \ \ \ 
 \del_m X^s \del_n X^s = \del_{(m}  \vp^a  \del_{n)} \bvp^a \ . }
The bosonic Lagrangian in  \one,\dbi\ has the following expansion
in powers of derivatives
$$
L = 1  + \ha \del_m X^s \del_m X^s  +  \four F^2  
 - \eight [ F^4 - \four (F^2)^2]
 - \ha (  F^2_{mn}    -  \four 
F^2 \d_{mn} ) \del_m X^s  \del_\n  X^s 
$$
\eqn\oopp{
+ \ \eight
[ (\del_m X^s \del_m X^s)^2  - 2 (\del_m X^s \del_n X^s) 
(\del_m X^u \del_n X^u)] + ...
\ , } 
where $F^2_{mn} = F_{mr}F_{nr} $,  
or, equivalently, 
$$
L = 1  + \ha \del_m \vp^a \del_m \bvp^a   + \four  F^2 
  -
 \eight [ F^4 - \four (F^2)^2]
 - \ha (  F^2_{mn}  - \four F^2 \d_{mn} ) \del_m \vp^a \del_n \bvp^a
$$
\eqn\yuy{
+ \ \eight
( \del_{m} \vp^a \del_{m} \bvp^a  
\del_{n}\vp^b \del_{n}\bvp^b   - \del_{m} \vp^a \del_{m} \vp^b  
\del_{n}\bvp^a \del_{n}\bvp^b - \del_{m} \vp^a \del_{m} \bvp^b  
\del_{n}\vp^b \del_{n}\bvp^a) 
\ . } 
The  knowledge of the bosonic action \yuy, the 4-derivative term 
in the $\N=2$ action \supeno\ and the condition of $SU(3)$ symmetry 
in  the chiral superfield sector allows one to deduce
the  analog of the term \supeno\ in the  
$\N=1$ superfield form of the $\N=4$ \BI action \liut.
This  gives the $\N=4$ generalization of the $I_4\sim F^4$ invariant 
in the \BI action \euci\ and its $\N=1$ ($W^2\bar W^2$) \solt\  and 
 $\N=2$ \supeno\ counterparts. 

Modulo the terms with $\del^2 \P$ the 
 only possible $SU(3)$ invariant  generalizations 
of the three $\P^4$  terms in \supeno\
  are
$$
P_1  = \ha (\bP^a \bP^b 
\del_m\P^a \del_m\P^b + \P^a \P^b\del_m\bP^a \del_m\bP^b ) 
\   ,  $$ 
\eqn\doo{
P_2  = \P^b \bP^b\del_m\P^a \del_m\bP^a \  ,  \ \  \ \ \ \ 
P_3  =  \P^b \bP^a \del_m\P^a \del_m\bP^b\  ,   }
\eqn\ide{
P_1 = - P_2 -P_3 +  \four \del^m \del_m ( \P^a \bP^a\P^b   \bP^b) 
}Using \eeeq\  
 and dropping the terms with $\del^2\vp $ which  can be eliminated by a field 
redefinition in the total  action containing $\P^2 + \P^4$ terms, 
 { as well as total derivative terms},  
 we find that the scalar field  parts of these invariants are 
\eqn\moreexp{
\int d^4\t \ P_1  = 2 \del_{m}\vp^a \del_{m}\vp^b  
\del_{n}\bvp^a \del_{n}\bvp^b \ ,  } $$
\int d^4\t \  P_2  = \del_{m} \vp^a \del_{m} \bvp^b  
\del_{n}\vp^b \del_{n}\bvp^a   - \del_{m} \vp^a \del_{m} \bvp^a  
\del_{n}\vp^b \del_{n}\bvp^b - \del_{m} \vp^a \del_{m} \vp^b  
\del_{n}\bvp^a \del_{n}\bvp^b\ ,   $$ $$ 
\int d^4\t \ \ P_3   =  \del_{m} \vp^a \del_{m} \bvp^a  
\del_{n}\vp^b \del_{n}\bvp^b - \del_{m} \vp^a \del_{m} \vp^b  
\del_{n}\bvp^a \del_{n}\bvp^b - \del_{m} \vp^a \del_{m} \bvp^b  
\del_{n}\vp^b \del_{n}\bvp^a \ . 
$$
To determine  the  relevant  
linear combination of  $P_1,P_2,P_3$  
 that generalizes the   $\P^4$ terms 
in \supeno\  we shall use the comparison of the scalar  field 
terms  with the corresponding $(\del X)^4$ structures  in the BI action
\yuy\ (which can be also  obtained 
 by dimensional reduction from the $F^4$ terms in the $D=10$ \BI action).
We 
find that the right combination is 
$P_1 + P_2 - 3 P_3$, i.e.  
the $SU(3)$ 
invariant generalization of \supeno\ to 
the case of several   chiral superfields is 
$$
{\bf I}_4 = W^{\a} W_{\a} \bar{W}_{\ad}  \bar{W}^{\ad} 
+ i (\bP^a \del_{m} \P^a - \P^a \del_{m} \bP^a) W^{\a} \s^{m}_{\a \ad} 
\bW^{\ad} $$
\eqn\fingen{
 +\  \ha ( \bP^a \bP^b 
\del_m\P^a \del_m\P^b+  \P^a \P^b\del_m\bP^a \del_m\bP^b) 
+  \P^b \bP^b \del_m\P^a \del_m \bP^a
-3  \bP^a \P^b \del_m\P^a \del_m\bP^b  \ . 
}
 Ignoring $\del^2 \P$ terms, the 
$\P^4$ terms in \fingen\  can be rewritten simply as 
$-4 P_3 + \four \del^\m \del_\m (  \P^a  \bP^a \P^b  \bP^b)$, or as  
\eqn\lop{
  2   \del_m (\P^a \bP^b) \del_\m ( \P^b \bP^a)
 - {\te { 3 \ov 4}}  \del^\m \del_\m (  \P^a \bP^a  \P^b \bP^b)
\ ,  } 
where the total derivative  term may be dropped in the action.
This    simple  $SU(3)$ invariant 
expression is the 
 generalization   of the $\N=2$  expression  \alta.

The $\N=4$ analog of the $F^4$ term in the \BI action
may be written also in terms of   $\N=2$  superfields
using projective superspace approach \refs{\gonr,\gonz}, 
or as  an integral of the 4-th power of the analytic on-shell
$\N=4$ superfield of \howest.

\newsec{Supersymmetric  Born-Infeld actions with `deformed' supersymmetry 
   from  D-brane
actions}
Component D3-brane   actions 
  with  $D=10$   space-time supersymmetry and 
local reparametrization invariance and  world-volume 
$\k$-symmetry 
  were constructed 
in
\refs{\ceder,\jhs,\bert} (in flat and generic curved backgrounds).

 The $\N=1, \ D=10$  supersymmetric \BI action  was  obtained in \jhs\ 
by fixing the static  gauge and a $\k$ symmetry gauge in the 
D9-brane in flat  type IIB  background. 
 Before 
 gauge fixing the action depends on  two $D=10$  Majorana-Weyl spinors 
$\t_1,\t_2$  and is invariant under two global $D=10$ supersymmetries. 
After $\k$ symmetry gauge fixing by setting $\t_2=0$ 
the action depends on the vector $A_\m$ and the  remaining  Majorana-Weyl spinor
$\t_1 \equiv \Psi$,\foot{The form of the action 
in  a different    $\kappa$-symmetry gauge --
Killing  gauge -- was  given in  \shmak.}
\eqn\biii{ S_{10} =  \int d^{10}  x 
    \sqrt {-\det (  \eta_{\m\n}  +  F_{\m\n}  -  
2 \bar \Psi \Gamma_\m\del_\n  \Psi 
 + \bar \Psi \Gamma^\rho \del_\m  \Psi 
\bar \Psi \Gamma_\rho \del_\n  \Psi)}\ . }
Here  $\Gamma_M$ are the $D=10$ Dirac matrices.
This action  is invariant under the  two original supersymmetries
supplemented by the $\k$ symmetry transformation 
to maintain the gauge. One  of the resulting symmetries
(corresponding to the spontaneously broken half of the  $\N=2$, $D=10$ supersymmetry)
 is 
realized non-linearly. Under the   other (unbroken 
 combination of the two original  
symmetries)  $\Psi$  has a  
homogeneous transformation law.
Similar action for the D3-brane is obtained by reduction to 4 dimensions \jhs\
(cf. \one) 
\eqn\bwii{  S_{4} =  \int d^{4}  x 
    \sqrt {-\det   ( \eta_{mn}  + \del_m X^s \del_n X^s  + 
 F_{mn}   + \psi_{mn} )
   } \ , } 
$$ \psi_{mn} \equiv
 2 \bar \Psi (\Gamma_m + \Gamma_s \del_m X^s ) \del_n  \Psi 
 + \bar \Psi \Gamma^s \del_m  \Psi 
\bar \Psi \Gamma_s \del_n  \Psi \ . $$
The unbroken supersymmetry of \biii\ or \bwii\ 
  has complicated form with terms of all orders 
in $F$ or in the fundamental scale parameter  (the inverse  string tension 
factor $2\pi \a'$ suppressed in \biii). It   may be thought of as an 
$\a'$-deformation of the linear supersymmetry transformations
of the $\N=1$, $D=10$  Maxwell multiplet. 
 This is  opposite \jhs\ to  what was the case 
 in the 
$\N=1, D=4$ superfield  action  discussed  in  section  4.1
where the  unbroken supersymmetry was  undeformed and was  simply
the original Maxwell supersymmetry while the broken supersymmetry
was non-linear (cf. \trans\ \bgg) and contained  terms of all orders 
in $F$ or in $\a'$. The  two 
 formulations  are presumably related by a field 
redefinition. 
 
The leading terms in the 
expansion of \biii\ (or its dimensional reduction \bwii)
 should be related by a field redefinition in the fermionic sector
to the  known $F^2 + \a'^2 F^4 $ deformation of the $\N=1$, $D=10$  Maxwell  action
 \refs{\bergr, \mera}. The latter 
 can be derived by a supersymmetric completion  of the bosonic 
gauge theory term  
 $F^4-\four (F^2)^2$   starting with the standard  the $\N=1, D=10$  SYM 
supersymmetry transformation laws and deforming the latter by $\a'^2$
corrections. Alternative route is  
 directly deducing the 4-fermion terms 
 from the open superstring amplitude
(the structure of the corresponding  invariant is dictated
by the standard massless mode superstring 4-point kinematic factor
\gsw).
In terms  of the $D=10$ gauge field strength $F_{\m\n}$ and 
$D=10$ Majorana-Weyl spinor $\Psi$
one finds (up to a field redefinition) the following 
supersymmetric completion of the $F^2+F^4$ terms ($2\pi \a'=1$) 
 \mera\foot{The corresponding 
non-abelian expression  \bergr\ is found by  taking the fields
to be $U(N)$ matrices and adding symmetrized trace.} 
 $$ 
L= - \four F^{\m\n} F_{\m\n} + {\te { i\ov 2}}
 \bar \Psi \G^\m\del_\m  \Psi 
 + {\te{ 1 \ov 8}} \bigg[
F_{\m\n}F^{\n\k}F_{\k\l}F^{\l\m} - \four  (F^{\m\n}F_{\m\n})^2
$$
 \eqn\comp{ 
+ \ha i \bar \Psi \G^{\m\n\k} \Psi  F_{\m\l} \del^\l F_{\n\k}
+ 2 i  \bar \Psi \G^{\m}\del_\n  \Psi  F_{\m\l}  F^{\l\n}
- {\te{ 1 \ov 3}} \bar \Psi \G^\m\del^\n \Psi \bar \Psi \G_\m\del_\n  \Psi 
\bigg] \ . }
Comparing the action \bwii\
to the actions  with   manifest linear $\N > 1$  supersymmetry 
discussed in sections 4.2, 4.3  we see that while 
\bwii\   depends only on the first 
derivatives of the scalars, the  manifestly supersymmetric 
actions  like \aact,\supeno,\fingen\
involve zero and second  derivatives of the scalars before one makes 
field redefinitions. Such field redefinitions  should   be 
 (at least partially) 
 responsible 
for  a non-linear modification of the supersymmetry transformation 
laws of the resulting translationally invariant actions like \bwii.

 Let us briefly mention that  one can also obtain  a 
 similar   action for a D3-brane moving in curved 
$AdS_5\times S^5$  vacuum background
of type IIB theory  \meets.\foot{
The  space-time  supersymmetric  and
$\kappa$-symmetric  D3-brane action was constructed   in terms of the
invariant  Cartan
one-forms defined on
the coset superspace $SU(2,2|4)/[SO(4,1)\otimes SO(5)]$.
The method used   is conceptually very close to the one
used in {\jhs} to find
the action of a  D3-brane propagating  in flat space
as  a $D=4$  `Born-Infeld plus Wess-Zumino' -type model on the
flat
 coset superspace
($D=10$ super Poincare group)/($D=10$ Lorentz group).} 
 The bosonic part of this action or the action  for  a D3-brane 
  moving 
near the core of another D3-brane  has the form 
\eqn\wzw{
S =   \int d^{4}  x \ 
    |X|^4 \bigg[ \sqrt {-\det (  \eta_{mn}  + 
Q |X|^{-4} \del_m X^s \del_n X^s  + 
 Q^{1/2} |X|^{-2} F_{mn} )} -1\bigg]  + S_{\rm CS}(X)  \ , } 
where $Q= 4 \pi N g_s \a'^2 $, \  $|X|^2 \equiv  X^s X^s$, \ $S_{\rm CS}
\sim  N \int_{5} \epsilon_{s_1 ...s_6}  \bar X^{s_1}  d\bar X^{s_2} \wedge ...\wedge d\bar X^{s_6}$, \  $\bar X^s \equiv X^s/|X|$. 
The supersymmetric extension of this action
generalizes \bwii.
This action 
should   coincide  with
the leading IR,   large $N$,  part
of  the {\it quantum}  $\N=4$  $SU(N)$ SYM effective action
obtained by keeping  the $U(1)$ $\N=4$  vector multiplet as an external
 background and
integrating out  massive SYM  fields
 \refs{\chep,\maldstr,\malda,\keski}.\foot{$S_{CS}$  which is $SO(6)$ invariant (does not depend
on $|X|$)  should have purely 1-loop origin.} 
One simple test of this conjecture is the following:
since  the quantum $\N=4$  SYM theory is conformally
invariant, the resulting action
should  also have (spontaneously broken by scalar field background and thus
non-linearly realized) conformal symmetry.
The non-linear conformal invariance  of the  bosonic part of the
static-gauge D3-brane
action  in $AdS_5$
background  was indeed demonstrated 
 in \refs{\malda,\kalo}.\foot{It was 
conjectured  in \malda\ (and demonstrated for the 
particular case when only the modulus of $X_s$ is 
non-constant) that  this non-linear 
symmetry may be fixing the structure of the action 
\wzw\ uniquely. This seems to be unlikely since superconformal
symmetry is 
not sufficient to restrict the form of the vector field 
 terms, and
the scalar terms should be related to
the gauge theory terms by supersymmetry.}
The validity of this ``quantum SYM $\to$ BI''  conjecture 
suggested  by the supergravity -- SYM correspondence
should rely on the   the existence of many 
 new non-renormalization theorems.

Like the flat-space action of \refs{\ceder,\jhs}, 
the action in \meets\ is 
invariant under the  32 global supersymmetries of
the \ads vacuum
 and $\kappa$-invariant.
Its conformal  invariance is a  consequence of the $SO(4,2)\times SO(6)$
isometry of the  \ads metric
and
 is manifest (linearly  realized) {before} the  static
gauge fixing.
It is only after  choosing the standard  Poincare coordinates
and fixing the static gauge and  appropriate $\kappa$-symmetry gauge
it    will    have a  ``SYM effective action'' interpretation
(details of  this procedure remain to be understood). 
Like the flat space action \bwii, it  will then  have 16 linear  and
16 non-linear (conformal)  supersymmetries, i.e. 
only the $ISO(3,1)\otimes SO(5)$  and 16 supersymmetries
 of the original symmetry will
remain manifest, but  the   superconformal symmetry
will be realized non-linearly.
While   for 
both D3-brane actions -- in flat space  and in \ads space --
 their gauge-fixed forms  have
only   16 linearly realized supersymmetries, 
the interpretation of the remaining 16 supersymmetries
as conformal  ones is possible only in the \ads case.\foot{This interpretation seems to depend on
a proper choice the $\kappa$-symmetry gauge which
should be different, e.g.,  from the $\theta_2=0$ choice
in \jhs.}
The  difference between the two actions
 is  related to the  fact that  while
the flat space action \bwii\ has explicit scale ($\sqrt{\alpha'}$),
the role of such scale in the \ads action
is played by the modulus of the scalar field.\foot{In contrast 
to the \ads one,
the flat-space \BI-type  D3-brane action
 is not, of course, 
 related to quantum SYM theory; instead,  the higher-order  terms in it
are  interpreted as  tree-level  string-theory $\alpha'$ corrections.}
As in the flat case, 
the resulting action should be invariant 
under complicated (`$X$-deformed') supersymmetry transformations.
Examples of similar  actions in  lower dimensions were constructed in 
\seth.

\newsec{Non-abelian generalization of \BI action}
The abelian \BI action represents the derivative-independent 
part of the open string tree level effective action.
In contrast, 
the  part of the 
string  effective action  for the non-abelian vector field
which depends on the field strength but not on its  covariant 
derivatives is not defined unambiguously since 
$[D_m,D_n] F_{kl} = [F_{mn}, F_{kl}]$. 
  One natural definition of the non-abelian \BI (NBI)  action
suggested in   \tsen\  which will be described below 
is based on  replacing $F_{mn}$ in the BI action by a 
non-abelain field strength and adding the symmetrized trace 
in front of the  $\sqrt{\det}$ action. 

\subsec{String theory  considerations  }
 This definition  can be motivated  from string theory as 
 follows \tsen.\foot{A  somewhat   different 
proposal was made  in \napp.}
One starts with the 
path integral representation  for the 
 generating functional for the 
   vector scattering amplitudes on the disc\foot{Here $P$ stands for 
the standard path ordering. As explained 
in \refs{\andrt,\andr}, the  $[A_m,A_n]$ term in $F_{mn}$ appears 
from   the  contact   terms
in the supersymmetric theta-functions  in the definition 
of the supersymmetric path ordering.}
 \eqn\acti { 
 Z(A) = < \tr\ P \exp i\int d \vp \ [ \dot x^m A_m (x) - \ha \psi^m\psi^n F_{mn}(x)] >  } $$
 = \int d^D x_0  < \tr\ P 
  \exp i \int d \vp \ [ \dot \x^m A_m (x_0 + \x) 
- \ha \psi^m\psi^n F_{mn}(x_0 + \xi)   ]> \  , $$
 where the trace is in the fundamental representation
of the Chan-Paton group, 
  $ x= x_0 + \x (\vp)$, \ 
 $0 < \vp \leq 2\pi$ and 
 the averaging is done with the free string propagator 
 restricted to the boundary of the disc, 
i.e.  with 
the action $\int (\x G\inv \x + \psi K\inv \psi ) $ 
($\vp_{12} \equiv \vp_1 - \vp_2$,  \ $\ep\to + 0$)
\eqn\tytr{
G(\vp_1,\vp_2) = {1\ov \pi}\sum^\infty_{n=1} {e^{-n\ep}
   \ov n} \cos n \vp_{12}   \ , \ \ \ \ \ \ \ \ 
 K(\vp_1,\vp_2) =   {1\ov \pi}\sum^\infty_{r=1/2 } {e^{-r\ep}
  } \sin r  \vp_{12}  \  .  }
Using the radial gauge $\x^m A_m (x_0 + \x) =0, \ A_m (x_0) =0$
(see, e.g., \shif)   one finds the following expansion 
in terms of symmetrized products of covariant derivatives of $F$
 at  $x_0$, 
 \eqn\rad{ 
 \int d\vp\ \dot \x^m A_m (x_0 + \x) =  \int d\vp\ \dot \x^m\big[
 \ha \x^n F_{nm} + 
  {\textstyle {1\ov 3}}\x^n \x^l D_l F_{nm} + 
   {\textstyle {1\ov 8}} \x^n \x^l \x^s D_{(s } D_{l)} F_{nm} + ... \big] \ .  }
Then 
 \eqn\actr { 
 Z(A) =  \int d^D x_0
  \big[ \LL (F)  + O( D_{(k} ... D_{l)} F) \big] \ ,  }
  \eqn\ttt{
 \LL (F) =  < \tr\ P 
  \exp\big[\ha { i} F_{nm}  \int  d \vp \ 
( \dot \x^m \x^n + \psi^m\psi^n) \big] >  \  .  }
Dropping all symmetrized covariant derivatives 
leaves us with $\LL(F)$. 
 The path integral in  \ttt\    is effectively non-gaussian
     because
 of the  path  ordering of the  $F_{nm}(x_0) (\dot \x^m \x^n)(\vp)$ 
 factors
 which is non-trivial if the matrices $F_{mn}$ do not commute. 
If we further {\it define} the NBI  Lagrangian  as part of $\LL(F)$ which does not 
contain commutators of $F$'s, or, more precisely, 
which is completely symmetric in all factors of $F$ in each monomial $\tr (F...F)$, we can then replace the trace 
in \ttt\ by {\it symmetrized} trace, i.e. treat $F_{mn}$ matrices as if they are 
commuting and thus drop the path ordering  symbol. 
 The resulting path integral is then 
 computable  exactly as in the abelian case  \ts\ 
\eqn\actrss { \LL(F)\  \to\ 
 L_{{\rm NBI}} (F) 
   = \Str <  
  \exp\big[{\ha i F_{nm} \int^{2\pi}_0  d \vp \ (
   \dot \x^m \x^n  +  \psi^m \psi^n )} \big] >   } $$ 
   = 
  \Str \big[ {-\det(\eta_{mn} + T\inv   F_{mn} ) }\big]^\n  \  ,  $$ 
\eqn\nnn{ \n = - \pi \int^{2\pi}_0  ( \dot G^2  -  K^2) 
=( - \sum^\infty_{n=1} e^{-2\ep n} +  \sum^\infty_{r=1/2 }
 e^{-2\ep r})_{\ep \to 0}  = \ha   \ ,  }
and thus 
\eqn\nonab{
L_{{\rm NBI}} (F) 
   = \Str \sqrt{- \det(\eta_{mn} +  T\inv  F_{mn} ) } \ . } 
This NBI action    represents in a sense
a  `minimal' non-abelian  extension of the abelian \BI action 
which is consistent with the  basic  requirement
of tree-level string theory -- overall  single trace 
of products of field strengths
 as matrices in the fundamental representation. 
Remarkably, it reproduces   exactly the   
$F^2 +\a'^2 F^4$ terms in the full non-abelian open superstring effective action\foot{The non-abelian $F^4$ terms were originally 
found in the $\Str$-form in \gross\ and in the equivalent 
$\tr$-form in \tsey.}
$$  
(2\pi\a')^{-2} \Str [\sqrt {-\det( \eta_{mn} 
+ 2\pi \a'   F_{mn})} -I]  $$  $$
=   \Str \big[\four   F^2_{mn}  -  {\te {1 \ov 8}} 
 (2\pi \a')^2
 \big(F^4 - \four (F^2)^2\big) + O(\a'^4) \big] $$
  \eqn\nbe{
 =  \tr \big[
\four F^2_{mn} -  {\te { 1 \ov 12}} (2\pi\a')^2 \big(F_{mn}F_{rn}F_{ml}F_{rl} +
\ha  F_{mn}F_{rn}F_{rl} F_{ml}  } $$ 
- \  \four F_{mn}F_{mn}F_{rl}F_{rl}
-{\textstyle {1\ov 8}} F_{mn}F_{rl}F_{mn}F_{rl} \big)
  + O(\a'^4) \big] \ .  $$
It should be stressed  that these $F^4$ terms represent  the full 
$O(\a'^2)$ term in the superstring effective action, 
i.e. all other possible 
 terms with covariant derivatives can  be redefined away at this order \tsey. 
 
In general, the full open string effective action 
is given by the sum of the three types of terms:
(i) NBI action \nonab; (ii)  $F^n$ terms containing  factors of 
 commutators of $F$'s; (iii)  terms with symmetrized 
covariant derivatives of $F$. 
While the separation of terms with symmetrized 
covariant derivatives is unambiguous, 
terms from (i) and (ii)  have similar $\tr (F...F)$ structure
and their sum reduces to the abelian  \BI action 
in the case  when $F$'s commute. 
It 
 is clear, of course,
that there is no reason to expect that the 
NBI action \nonab\  should reproduce the  full 
string  theory expression at higher than $\a'^2$ orders
(i.e. $\a'^4 F^6+...$).\foot{Therefore,  
 possible disagreements  with  predictions of the full string theory
effective action like  the one observed  in \hasht\
(where  quadratic fluctuations  in a constant 
abelian $F_{mn}$ background  were discussed both from NBI 
action and string theory points of view) 
should not be  unexpected.} 
Still, the symmetrized trace  action 
has several  exceptional  features   and may indeed 
provide  a good approximation to string (or D-brane) 
dynamics in certain  situations, e.g., described by   nearly 
commuting  or nearly 
covariantly constant field strengths,   
or by BPS configurations.

\subsec{Properties of the  symmetrized trace action}

Before discussing some properties and generalizations of the 
NBI action \nonab\ let us make its    definition more explicit.
Expanding the abelian \BI Lagrangian in powers of $F$ we may 
define the Lorentz tensors $C^{m_1n_1... m_{2k} n_{2k}} $ 
as the coefficients in\foot{Since $\det(\eta + F) = \det (\eta + F^T) = \det (\eta -F)$
the expansion of BI  and thus of NBI action contains only even powers
of $F$.}
\eqn\expa{
\sqrt{-\det(\eta_{mn}  + F_{mn})}  = \sum_{k=0}^{\infty} C^{m_1n_1... m_{2k} n_{2k}}
 F_{m_1n_1} ... F_{m_{2k} n_{2k}} \ . }
If $F_{mn} = F^a_{mn} T_a$  where $T_a$ are  
generators of the gauge group
(in the  fundamental representation) 
the non-abelian \BI action is defined by 
\eqn\expan{
\Str \sqrt{-\det (\eta_{mn} I  + F_{mn})}  \equiv
 \sum_{k=0}^{\infty} d_{a_1 ...a_{2k}} C^{m_1n_1... m_{2k} n_{2k}}   
 F^{a_1}_{m_1n_1} ... F^{a_{2k}}_{m_{2k} n_{2k}} \ .  }
Here the totally symmetric tensors
\eqn\ddd{
d_{a_1 ...a_p} = \Str( T_{a_1} ... T_{a_p}) \equiv
{ 1 \ov p!} \tr ( T_{a_1} ... T_{a_p} + {\rm all \ permutations}) \  }
 are the (adjoint action) invariant   tensors of the gauge Lie  algebra
($\sum_i d_{a_1 ... a'_i ... a_p} f^{a'_i}_{ \ a_i b} =0, \ 
\ [T_a,T_b] = f^c_{\ a b} T_c$).\foot{In general, for $SU(N)$ there are $N-1$(=rank) 
basic or primitive tensors ${d}_{a_1...a_r}$  in terms of which all other
$d_{a_1...a_n}$  can be expressed (see, e.g., \moun). 
The primitive symmetric tensors  define 
the Casimir operators $I_r = {d}_{a_1...a_r} T^{a_1} ...T^{a_r}$.}
The definition \expan\ is thus quite  natural from  the 
 mathematical  point of view.

For example, for $SU(2)$ one has 
$T_a= \s_a, \  T_{a} T_{b} = \d_{ab}  + i \ep_{abc} T_c$,  so that   $  
d_{a_1...a_{2n}} = 2 \d_{(a_1a_2} ...\d_{a_{2n-1,2n})}$.  
 For $SU(3)$   all 
$d_{a_1...a_{2n}}$ are expressed in terms of $d_{ab} \sim  \d_{ab}$ 
and $d_{abc}$.  
The simple structure of $d_{a_1...a_{2n}}$ in the $SU(2)$ case allows one 
to write down the $SU(2)$ NBI Lagrangian in the following form:
$L_{{\rm NBI}}{(SU(2))} = < \sqrt{\det(\d_{mn}  + {\cal F}_{mn})} >, $
where ${\cal F}_{mn} = t_a F^a_{mn}$ and the averaging is done 
over the free   gaussian   variable $t_a$ with the rule
$< t_a t_b > = 2 \d_{ab}$, i.e. 
$ <...> \sim  \int [dt] \exp (- t_a t_a) ...$.\foot{Similar representation might exist 
for  higher $SU(N)$ groups if  additional primitive  invariant tensors 
 are added as  coupling  constants  to the action for $t_a$ 
(e.g., $d_{abc} t_a t_b t_c + ...$), making it non-gaussian.}

The fact that
under $\Str$   one can effectively  treat 
 the factors of $F$ as commuting    simplifies
the analysis  of the consequences of the  NBI action
 (see \refs{\hasht,\hashi}). 
Indeed, most of  the properties of the abelian \BI action  have direct   non-abelian analogs in the NBI action  case.
In particular,  one can show that:
   (i) covariantly constant fields $D_m F_{kl}=0$ are solutions of the NBI
equations;\foot{The variation of the NBI action is 
$\Str[\sqrt{-\det H_{kl} } (H^{-1})^{ mn}  (D_m \delta A_n -
 D_n \delta A_m) ]$, where
$H_{mn} \equiv \eta_{mn} +  F_{mn}$. 
Since $\Str$ \ddd\ is the sum of terms with ordinary traces
which have the usial cyclic symmetry, one can always put the term 
with variation to the right of all others. In general,
for a set of matrices $M_i,K$ one has 
$\Str (M_1...M_n K) = \tr (M_{(1}...M_{n)} K)$ where
$M_{(1}...M_{n)}\equiv { 1 \ov n!} \sum (M_1...M_n+$all permutations).  }
\ (ii) the NBI action has the same BPS solutions
 (waves, instantons, monopoles, etc.)
as the  YM action $\tr F^2_{mn}$
 \refs{\hashi}.\foot{See also \nun\
for a discussion of Bogomol'nyi 
 relations  in the abelian 
$\N=1$ \BI theory \solt\  combined with  a  Higgs scalar  Lagrangian
(such  theory may result as a certain  approximation from  the 
NBI action). Monopole solutions in the  non-abelian theory combined with Higgs 
sector were considered in \gon.}
 The NBI  action and equations of motion 
reduce\foot{As in the abelian case \euc, on 
 self-dual configuration $\det(\d_{mn} +  F_{mn})$ 
reduces to  a perfect square 
$[1 + \four (F_{mn})^2]^2 $.}
 on such configurations 
simply to the YM  action and the  YM equations $D_m F^{mn}=0$.
As in the abelian case \foo,\euc\ one can 
show that in four (euclidean) dimensions
  $$ L_{\rm NBI} = \Str \sqrt{{\rm det}_4 (\d_{mn}  + F_{mn})} 
=  \Str \sqrt { 1 + \ha F_{mn}F^{mn} + 
{\textstyle {1\ov 16} } (F_{mn}\sF^{mn})^2}  $$
\eqn\foou{
= \ \Str  \sqrt{(1 + \four F_{mn}\sF^{mn})^2 + \four (F_{mn} -\sF_{mn})^2}\ . }
However, this formal representation does not mean that 
the resulting action  is expressed in terms 
of the two Lorentz scalars only: $\Str$ 
 includes all possible orderings of the $F_{mn}$ factors.\foot{Thus the 
existence of 
a BPS bound similar to \euc\
is  not obvious   \gibbo\   (cf. \brec). 
The   inequality  relation  
for  $\Str\sqrt{\det(...)} $ action  needs, in general, 
 a separate proof  different from  the abelian argument
(note also  that $\Str$-action \expan\ 
 is defined  perturbatively and its  global properties depend 
on convergence of the series).
The BPS bound seems to  hold at least  in the $SU(2)$ case
as can be seen  from the `abelian' 
representation $L_{{\rm NBI}}{(SU(2))} = < \sqrt{\det(\d_{mn}  + {\cal F}_{mn})} > \  $
mentioned above (the averaging is done with a positive definite measure). 
}

The bosonic NBI action admits straightforward  supersymmetric 
extensions generalizing the abelian actions like 
\solt\  (see \gonor)  or \biii.
The 
non-abelian generalization of \comp\ or, equivalently, 
the supersymmetric
 version of the $\tr (F^2 + F^4)$  terms in \nbe\ 
 invariant under  $\a'$-deformed supersymmetry 
was found in \bergr. 

\subsec{ Non-abelian D-brane actions }
The  non-abelian  actions  for clusters 
of D-branes can be  again obtained by dimensional reduction
of the NBI action \tsen\  (cf. \qer). 
As  was argued in \witt, for a   system of 
$N$ parallel D-branes the fields 
$(A_m, X_s)$ become $U(N)$ matrices and the  low-energy 
  Maxwell action  is generalized to the 
$D=10$  Yang-Mills action reduced to $p+1$ 
dimensions. The full action  including higher order terms 
is determined by   the 
dimensional reduction
of the   open string effective action with the gauge potential 
components replaced by the matrix-valued fields $A_\m=(A_m, A_s= T X_s)$.
This follows directly   from the  non-abelian 
generalization of the  partition function approach to the derivation of
D-brane actions  discussed in \tse.
T-duality  relates 
the Neumann $A_s$ and Dirichlet $A_a$ vertices 
in the exponent in
 \eqn\tytro{Z= < \tr\ P \exp i\int d\vp [\del_\vp x^m A_m(x) + 
 \del_\bot x^s A_s(x)] > \ . }
In view  of the above discussion, 
 in  situations 
when covariant derivatives and their commutators are small, 
it is natural to assume  that  the  most important part of these
corrections   is  represented
 by the NBI action \nonab. Then  we arrive at 
the 
following non-abelian generalization  
 of \qer\ \tsen\
$$S_p = \T_p \int d^{p+1} x \
 \Str \sqrt {-\det 
(\eta_{\m\n}  +   T\inv F_{\m\n})}  $$
 \eqn\rrer{= \ 
  \T_p   \int d^{p+1} x \
 \Str  \bigg[  
    \sqrt {-\det {\cal G}_{mn}} \sqrt {\det {\cal G}_{rs}} \bigg] \ , }
\eqn\meew{
{\cal G}_{mn} \equiv  \eta_{mn} + {\cal G}^{rs}(X)  D_m X_r D_n X_s  
+ T\inv F_{mn} \ , \ \ \ \ \ 
{\cal G}_{rs}(X)  \equiv \d_{rs} - i T [X_r, X_s] \ . }
Here $\Str$ applies   not to individual $A_m$ and $X_s$ 
but 
 to the products of 
 components of the {\it field strength}  $F_{\m\n}$, i.e.
$F_{rs} = -iT^2 [X_r,X_s]\ , 
 \ F_{mr} = T D_m X_r= T(\del_m X_r - i [A_m,X_r])$ and $F_{mn}(A)
= \del_m A_n - \del_n A_m - i [ A_m, A_n]$.
As in \qer, we have used   the simple
determinant identity (which is  applicable  under $\Str$). Compared to the abelian case, now 
there is a non-trivial  extra factor of 
the  determinant of the `internal metric' 
 ${\cal G}_{rs}(X)  \equiv \d_{rs} - i T [X_r, X_s]$
(which is equal to 1 if $X_s$ commute).\foot{Some quantum corrections
in such non-abelian D-brane actions were discussed in \odint.}

There should exist a non-trivial generalization
of the  action \rrer\ to the case 
 when D-branes are  put in a curved
background.  One natural suggestion is to consider 
background fields  as  functions of $X_s= x_s I + X'_s$
where $x_s$ is  the center of mass coordinate  and expand in power
 series in  $SU(N)$ coordinates $X'_s$. For example, for the dilaton interaction 
$\tr (\phi F^2_{mn})$ this  leads to $\sum { 1 \ov n!}  \del_{s_1}
.. \del_{s_n}  \p (x) \tr (X'_{s_1}...X'_{s_n} F^2_{mn})$  \refs{\kleb,
\klebb}.
As was pointed out  in  \wiit,   
to get SYM operators 
 from short multiplets of $\N=4$  superconformal algebra 
one should  symmetrize  products of $X$'s
 (or of their $\N=1$ superfield 
counterparts).  This prescription is also   supported 
and  made more universal by the  analysis 
of  the  supersymmetric versions of the above  higher momentum dilaton 
operators in 
\ktv.   Note  also that   powers
 of $X$ and $F$  are related 
by linearized $\N=4$ supersymmetry
(in particular, $\Str(XXXX) $ is related to 
$\Str(FFFF)$  \refs{\ferr,\aha}).
This suggests that in the  general 
external background case  
the symmetrized trace in the NBI action should apply 
to both   the components of 10-d field strength 
$F_{mn}, D_m X_s, [X_r,X_s]$ {\it and}   powers 
of $X_s$ in the Taylor expansion
of the background fields.

\newsec{Derivative corrections to 
 \BI action in open superstring theory }
The  field strength derivative corrections to the leading \BI part of the 
open string effective action  can be computed   using either 
S-matrix  or sigma model (path integral  or beta-function)   approach.
The advantage of the latter is that though it is   
based on the  expansion  in derivatives  $\del_k F_{mn}$, in the abelian case 
it is still 
non-perturbative in powers of $F_{mn}$.
As  explained  in \refs{\ts,\andr}, one may  start with 
the path integral expression \oct\ or \acti,\rad, 
expand the boundary  action   in powers of $\del F$ and compute  
the resulting correlators with the $F_{mn}$-dependent 
(super)propagator on the disc 
which in the superstring case has the form \refs{\abo,\ts,\andr} 
(cf. \tytr)
$$\hat G_{mn} =  G_{mn} (\vp_1,\vp_2|F) - \t_1 \t_2  K_{mn} (\vp_1,\vp_2|F) $$
$$=\  { 1 \ov \pi} \sum^\infty_{n=1} {e^{-\ep n}\ov n} \big( f_{mn} \cos n \vp_{12} - i h_{mn} \sin n \vp_{12} \big) 
$$
\eqn\proo{- \ {i\ov \pi}  \t_1 \t_2 
 \sum^\infty_{r=1/2 } e^{-\ep r} \big( h_{mn} \cos r \vp_{12} - i f_{mn} \sin r\vp_{12} \big) \ , }
$$ f_{mn} \equiv [(\eta -F^2)^{-1}]_{mn} \ , 
 \ \ \ \ \ \   
h_{mn} \equiv  F_{mk} f_{kn} =   [F(\eta -F^2)^{-1}]_{mn} \ ,  $$
where we have absorbed $2\pi \a'$ into the matrix $F_{mn}$
and $(\eta -F^2)_{mn} \equiv  \eta_{mn} - F_{mk}\eta^{kl}  F_{ln}$.  
After the renormalization of the partition function $Z(A)$, i.e.
eliminating    logarithmic divergences   by a redefinition 
 of the vector potential,\foot{This 
 corresponds to subtraction of massless exchanges 
in the string amplitudes. Renormalization scheme
ambiguity corresponds to  field redefinition ambiguity
in the S-matrix approach \refs{\ts}.}
one finds \andr\ that the resulting effective superstring 
 Lagrangian  has the 
following structure (we  again 
suppress the  factors of $2\pi \a'$  which multiply
each  $F_{mn}$  and each pair of derivatives) 
\eqn\stru{
L= \sqrt {- \det (\eta_{mn} +  F_{mn})}\bigg[
1 + {\cal F}^{klmn abcd}(F) \del_k \del_l F_{mn} \del_a \del_b F_{cd} 
+ O (\del^6) \bigg] \ , }
where the function 
${\cal F}(F) \sim   F^2 +  F^4 + ... $ 
can be, in principle, computed  exactly.
 The leading $F^2$ term in $\cal F$  is easy to find 
  by comparing  this action with  the momentum expansion 
of the standard expression \gsh\ for 
the 4-vector superstring amplitude on the disc \andr\ 
$$
L= \sqrt {- \det {(\d_{mn} +  F_{mn})}}
$$
$$ - {\te{ 1 \ov 96} } 
\bigg( \del_a\del_b  F_{mn} \del_a\del_b  F_{nl} F_{lr} F_{rm} 
  + \ha \del_a\del_b  F_{mn}   F_{nl} \del_a\del_b F_{lr} F_{rm}
$$
\eqn\deri{- \four \del_a\del_b  F_{mn}   F_{mn} \del_a\del_b F_{lr} F_{lr}
- {\textstyle {1 \ov 8}} \del_a\del_b  F_{mn} \del_a\del_b  F_{mn}  F_{lr} F_{lr} \bigg)
\  + O( \del^4 F^6 ) \ . }
The $FF \del \del F \del  \del F$ term  in \deri\ 
defines a  super-invariant  which appears also in other contexts (e.g., 
as a divergent 1-loop correction 
to quantized  supersymmetric  \BI action \refs{\shmak,\zan}). 
Starting with \deri\ in 10 dimensions and applying 
dimensional reduction  as in \qer,\rrer\ one may determine 
the corresponding higher-derivative string $\a'$  
corrections 
to  D-brane actions.\foot{Note that the 
 $O(\del^2 X)$  higher-derivative  terms 
   discussed in \peri\  can be redefined  away \roch: 
the scalar terms in \peri\ 
are related by field redefinition and integration by parts 
 to $(\del X)^4/X^4$ 
terms which accompany $F^4/X^4 $ terms (cf. \oopp) 
in the $\N=4$ SYM  
1-loop effective action. 
}

For comparison, in the bosonic string case 
one finds \refs{\ts,\andr}
$$L= \sqrt {- \det (\eta_{mn} +  F_{mn})}\bigg[
1 + {\cal F}^{kmn acd}(F) \del_k  F_{mn} \del_a  F_{cd} 
+ O (\del^4F^k) \bigg] 
$$
\eqn\otru{= \ \sqrt {- \det (\eta_{mn} +  F_{mn})}  }
$$-\ {\te { 1 \ov 48 \pi}} \bigg(
   F_{kl}F_{kl} \del_a F_{mn} \del_a F_{mn} 
+ 8  F_{kl}F_{lm} \del_a F_{mn} \del_a F_{nk}
- 4  F_{la}F_{lb} \del_a F_{mn} \del_b F_{mn}\bigg)
+ O(  \del^2 F^6) ,  $$
where the leading $F^2$ term in the  function 
${\cal F}(F)  $  can be found from the string 4-point amplitude \andr\ 
or from the 2-loop beta-function computation \andre.
 

\newsec{Acknowledgements}
I am  grateful to I. Buchbinder, F. Gonzalez-Rey,  I. Klebanov,  H. Liu   and 
  R. Metsaev
for useful discussions of related  questions.
This work was supported  in part  by
 the PPARC,  
 the EC TMR programme ERBFMRX-CT96-0045,  
  INTAS grant No.96-538,  and NATO  Collaborative Linkage Grant 
  PST.CLG 974965. 

\appendix{A}{
 Born-Infeld  action from string partition function on a disc}

Let us  review the original computation \frts\ 
of the partition function \oct\  in the abelian $F_{mn}=\const$ 
background  in a slightly generalized  form:  we shall assume that 
the   boundary part of the string action  contains  also  the 
usual  `particle' term  $(\dot x^m)^2$.  The full (Euclidean)
 string action 
on the disc is  then  
\eqn\ful{
I= \int d^2 \s\ \ha T \del^a x^m \del_a x^m  + 
 \int^{2\pi}_0  d \vp\  [ \ha M_{mn} \dot x^m \dot x^n   - 
    i  \dot x^m A_m (x) ] \ . }
Here  $M_{mn}$ may be interpreted as  a condensate
of the open string  massive mode; taken at  an `off-shell' 
value $M_{mn}=\const$ this term breaks conformal invariance 
of the sigma model. In what follows we shall set $M_{mn} = M \delta_{mn}$
and treat this term as a formal `regularization' of the boundary kinetic operator. Integrating over the values of  the string coordinate at internal points
 of the disc  we arrive at the following effective action at the boundary of the disc (we  isolate the constant zero mode $x^m_0 = x^m-\xi^m$)
\eqn\fulb{
I_{\rm bndry}= 
\ha  \int^{2\pi}_0  d \vp\  [  T  \x^m G^{-1}  \x^m   + 
 M  \dot \x^m \dot \x^m   +   i F_{mn} \x^n \dot \x^m   ] \ ,  }
where $\x^m = \sum^\infty_{n=1} (a^m {\rm cos}\ n \vp 
 + b^m {\rm sin}\ n \vp)$ 
and  the scale-invariant (`first order') non-local operator $G^{-1}$ is the 
inverse of the  restriction of the Green function on the 
disc to its boundary, 
\eqn\tytre{
G(\vp_1,\vp_2) = {1\ov \pi}\sum^\infty_{n=1} 
{1  \ov n} \cos n \vp_{12}   \ , \ \ \ \ \ \ \ \ 
G\inv (\vp_1,\vp_2) = {1\ov \pi}\sum^\infty_{n=1} 
 { n} \cos n \vp_{12}   \ .} 
The action \fulb\ thus contains both effectively first-order
($\sim T$)  and second-order ($\sim M$) 
derivative parts and  interpolates between the 
string-theory case $T\not=0, \ M=0 $ which was
 discussed in \frts\  
  and the  standard particle  case   $T=0, \ M\not=0$
which  appeared in the 
Schwinger computation of  $\log \det (-D^2(A))$. 
The resulting partition function will  interpolate between the \BI 
(``$\sqrt{ 1 + \f^2}$'') and 
Schwinger (``$ \f\ov {\sinh \f}$'') results. 

Putting $F_{mn}$ in the  block-diagonal form and concentrating on the first 
 $(1,2)$ block  
we find, integrating over the  coordinates $\x^1,\x^2$ as in \frts\  ($F_{12}= \f$):
$$ Z_{12}  =    Z_{12} (M) Z_{12} (F,M)  \ , \ \ \ \ \  $$
\eqn\ytre{ Z_{12} (M) \sim \prod_{n=1}^\infty (Tn + M n^2)^{-2}
\sim\   M \ [\prod_{n=1}^\infty (1 + { TM\inv \ov n} )]^{-2} \ , } 
\eqn\twq{Z_{12} (F,M) = 
\prod_{n=1}^\infty [  1 + {\f^2 \ov (T + M n)^2 }  ]^{-1} \ .  }
$Z_{12} (F,M)$  depends only on the ratios
$T\inv \f$ and $T\inv M$. 
We shall ignore  the (divergent) $F$-independent  factor   $Z_{12} (M)$
which can be absorbed into the renormalization
of tachyon  coupling  at the boundary.
When $T=0$  the  factors in the product in $Z_{12} (F,M)$ are
 $n$-independent, and  using the  regularization prescription 
$\prod_{n=1}^\infty c =  c^{-1/2}$  as in \frts\ 
 (with linear divergence again absorbed into the  tachyon  coupling \ts) we get the \BI result $Z_{12}(M=0, F)  =\sqrt{ 1 + (T\inv\f)^2}$. 
When $T=0$ we get the  Schwinger result
$Z_{12} = { \pi M\inv \f \ov \sinh (\pi M\inv \f)}$.\foot{In the
 1-loop field theory computation context 
$M= \pi/s^2$ where $s$ is the proper time  
parameter which one is still to integrate over.}
In general, for $M\not=0,$ \   $Z_{12} (F,M)$  
 is `more convergent' than for $M=0$ 
(i.e. $M$ plays the role of an effective regularization parameter)
and  
is given by a  combination
 of $\Gamma$ functions
\eqn\gaaa{
Z_{12} (F,M) = { \G ( { T + M +  i\f \ov M}) 
  \   \G ( { T + M -  i\f \ov M})  \over
[ \G ({ T + M   \ov M}) ]^2 }
 \ . } 
In the case of the electric field background  ($i\f \to E$) 
 the  partition function \gaaa\  becomes
\eqn\eeerP{
Z_{12} (E,M)
=   { \G ( { T + M + E \ov M}  )
  \G ( { T +M -  E \ov M})  \over
[ \G ({ T + M   \ov M}) ]^2 }
 \ . } 
This partition function   is well-defined for $|E| <   T + M $, 
which is a  generalization  of the critical field strength condition
($|E|  < T$) 
for the \BI action.

The general expression for the partition function is thus 
given by the product of factors for each eigenvalue $\f_p$  of the field 
strength
\eqn\aaa{
Z (F,M) = \prod^{D/2}_{p=1} \bigg[
  { \G ( { T + M + i\f_p \ov M} )\ 
  \G ( { T + M -  i\f_p \ov M})  \over
[ \G ({ T + M   \ov M}) ]^2 } \bigg] 
 \ . } 
It is easy to see that indeed
\eqn\uytr{Z(F,M)|_{T\to 0} \to \prod^{D/2}_{p=1} { \pi  M\inv
 \f_p \ov \sinh ( \pi  M\inv \f_p )} \ .   }
At the same time, 
taking  here the limit $M\to 0$ and 
using the  Stirling formula 
$\G({z \to \infty})  = \sqrt{ 2\pi \ov z}  ({z\ov e})^z [ 1 + O({1\ov z})]$
we  find that 
\eqn\liim{
Z (F,M\to 0 ) = \prod^{D/2}_{p=1} 
  \sqrt{ 1 + (T\inv \f_p)^2} \ [1 + (T\inv \f_p)^2]^{T\ov M} 
({ 1 + iT\inv \f_p \ov   1 -  iT\inv \f_p   })^{i\f_p\ov M}\  [ 1 +  O(M) ] \ . 
 } 
$M$ thus plays here the  role  of a   cutoff.
Eq. \liim\  reduces to the \BI expression after  renormalizing 
 the   divergent term in the exponent 
\eqn\yutr{Z (F,M \to 0) =  \sqrt{ \det(\d_{mn} +  T\inv F_{mn})} \ \ e^{ { 1 \ov M } f(F)} \ . }
The linearly divergent term  cancels out in the  superstring case
as in  \nnn.

The superstring generalization  of  the action 
\fulb\  contains  three extra fermionic  terms: \ 
\eqn\tyui{\ha \int d\vp ( T \psi^m K\inv \psi^m + M \psi^m \dot \psi^m 
 +  i F_{mn} \psi^m \psi^n)  \ , }
where $K$ is defined in \tytr. 
As a result, \twq\ is replaced by
\eqn\twff{
Z_{12} (F,M) = 
{ { \prod_{r=1/2}^\infty [  1 + {\f^2 \ov (T + M r)^2 }  ]  } 
\over  { \prod_{n=1}^\infty \ [  1 + {\f^2 \ov (T + M n)^2 }  ] } } 
 = \prod^\infty_{n=1}  \bigg[  { 1 + {\f^2 \ov (T -  { 1 \ov 2}  M  +  M n)^2 } \over
                 1 + {\f^2 \ov (T   +  M n)^2 }    }           \bigg] \ .  }
Then the  generalization  of \aaa\ becomes ($D=10$)
\eqn\faa{
Z (F,M) = \prod^{D/2}_{p=1} \bigg[
  { \G ( 1 +  { T + i\f_p \ov M} )\ 
  \G (1 +  { T  -  i\f_p \ov M})  \over
\G ( { 1\ov 2}  +  { T + i\f_p \ov M} )\ 
  \G ( { 1\ov 2}  +  { T  -  i\f_p \ov M}) } 
{ [\G ({ 1\ov 2}  + { T   \ov M}) ]^2 \ov
[\G ({1}  + { T   \ov M}) ]^2 }    \bigg]
 \ . } 
This expression is regular in the $M\to 0$ limit and reduces simply to 
the  \BI action\foot{The fermionic factors give 
 only  divergent contributions ($ [\Gamma ( \ha + z)]_{z\to\infty}
 \to \sqrt { 2 \pi \ov z}
({z \ov e})^{ {1\ov 2} + z} [ 1 + O( { 1 \ov z}) ] \to   z^z $) 
that cancel the $1\ov M$ divergences coming from 
the bosonic sector.} 
\eqn\poi{Z (F,M \to 0) =  \sqrt{ \det(\d_{mn} +  T\inv F_{mn})} 
 \ [ 1 + O({  M}) ] \ .  }
Thus $M$ plays indeed the role of a natural regularization
parameter. 
In the  limit $T\to 0$ one finds
$Z (F,M) = \prod^{D/2}_{p=1} { \pi M\inv \f_p \over \tanh (\pi M\inv \f_p)}$.

Similar expression for the partition function 
can be found using the Green-Schwarz light-cone gauge   string action
as in \tsey.
Assuming that the vector field strength
 has only spatial (magnetic) components $F_{ij}$ one 
is to replace \tyui\ by 
$\ha \int d\vp ( T S^a K\inv S^a + M S^a \dot S^a 
 +  i \hat F_{ab} S^a S^b ),  $
where $S^a$  is  an  $SO(8)$ spinor,  $\hat F_{ab} = { 1 \ov 4} (\gamma^{ij})_{ab} F_{ij}$ and $K$  is again given by \tytr\
($S^a$ is the restriction of  the l.c. GS 
spinor variable  which is a  2-d spinor 
to the boundary of the disc and  thus  like $\psi^m$ is 
antiperiodic in $\vp$  \mrt).
The resulting partition  function  has the same form as \faa\ (with $D=8$, i.e.
$p=1,2,3,4$)
but with $\f_p$ in  the fermionic contributions (two $\G$-function factors
in  the denominator) replaced by 
the eigen-values $\hat \f_p$  of the matrix $\hat F_{ab}$,
\ $\hat \f_1 = \ha (-\f_1 + \f_2 + \f_3 + \f_4)$, etc.
(see \ttts\ for  a discussion of a similar  partition function on
 the annulus).
The $M\to 0$  limit of the resulting partition function  has 
 again the  \BI  Lagrangian as its finite  factor, 
but  as in the bosonic case  \yutr\ 
and 
 in contrast  to the NSR case \poi\ the divergent $1\ov M$ terms 
here do not cancel, i.e.  in this 
case the $M$-regulator does not seem to 
preserve the world-volume  supersymmetry
(the  same conclusion
is reached in the case of the exponential regulator
used  in  \tytr).  At the same time, 
 the  formal $\zeta$-function regularization
implies \refs{\frts,\mrt}  that the bosonic factor is the BI one while
  the  fermionic contribution is simply  equal
to 1 ($\zeta(0, 1)= \zeta(0) = - \ha, \ \zeta(0, \ha ) =0$).

\listrefs
\bye